\documentclass{aa}  

\usepackage{xcolor}
\usepackage{graphicx}
\usepackage{txfonts}
\begin{document}

   \title{Temporal evolution and correlations of optical activity indicators measured in Sun-as-a-star
   observations}

   \subtitle{}

   \author{J. Maldonado\inst{1}
        \and
        D. F. Phillips \inst{2}
        \and
	X. Dumusque \inst{3}
        \and
        A. Collier Cameron \inst{4,5} 
        \and
        R. D. Haywood \inst{2,18}  
        \and A. F. Lanza \inst{6}
	\and G. Micela \inst{1}
        \and A. Mortier \inst{7}
        \and S. H. Saar \inst{2}
        \and A. Sozzetti \inst{8} 
        \and K. Rice \inst{9,10}
	\and T. Milbourne\inst{11,2}
        \and M. Cecconi \inst{12}
        \and H. M. Cegla \inst{3,19}
        \and R. Cosentino \inst{12}
        \and J. Costes \inst{13}
        \and A. Ghedina \inst{12}
        \and M. Gonzalez \inst{12}
        \and J. Guerra \inst{12}
	\and N. Hern\'andez \inst{12}
        \and C.-H. Li \inst{2}
        \and M. Lodi \inst{12}
        \and L. Malavolta \inst{6} 
        \and E. Molinari \inst{14} 
        \and F. Pepe \inst{3}
        \and G. Piotto \inst{15,16} 
        \and E. Poretti \inst{12,17}
        \and D. Sasselov \inst{2}
	\and J. San Juan \inst{12}
        \and S. Thompson \inst{7}
        \and S. Udry \inst{3}
        \and C. Watson \inst{13}
}

  \institute{INAF - Osservatorio Astronomico di Palermo, Piazza del Parlamento 1, 90134 Palermo, Italy\\
	     \email{jesus.maldonado@inaf.it}
             \and
             Harvard-Smithsonian Center for Astrophysics, 60 Garden Street, Cambridge, MA 02138, USA  
             \and
              Observatoire Astronomique de l'Universit\'e de Gen\`{e}ve, 51 Chemin des Maillettes, 1290 Sauverny, Suisse 
             \and
             SUPA, School of Physics and Astronomy, University of St Andrews, North Haugh, St Andrews KY16 9SS, UK 
	     \and
	     Centre for Exoplanet Science, University of St Andrews, St Andrews, UK  
             \and
	     INAF - Osservatorio Astrofisico di Catania, Via S. Sofia 78, 95123 Catania, Italy 
             \and
             Astrophysics Group, Cavendish Laboratory, University of Cambridge, J.J. Thomson Avenue, Cambridge CB3 0HE, UK
	     \and
	     INAF - Osservatorio Astrofisico di Torino, via Osservatorio 20, 10025 Pino Torinese, Italy 
             \and 
	     SUPA, Institute for Astronomy, Royal Observatory, University of Edinburgh, Blackford Hill, Edinburgh EH93HJ, UK
	     \and 
	     Centre for Exoplanet Science, University of Edinburgh, Edinburgh, UK
             \and
             Department of Physics, Harvard University, 17 Oxford Street, Cambridge MA 02138, USA
             \and
            INAF - Fundaci\'on Galileo Galilei, Rambla Jos\'e Ana Fernandez P\'erez 7, E-38712 Bre\~{n}a Baja, Tenerife, Spain 
             \and 
	     Astrophysics Research Centre, School of Mathematics and Physics, Queen's University Belfast, University Road, Belfast, BT7 1NN, UK 
             \and 
	     INAF - Osservatorio Astronomico di Cagliari, via della Scienza 5, 09047, Selargius, Italy 
             \and
             INAF - Osservatorio Astronomico di Padova, Vicolo dell'Osservatorio 5, 35122 Padova, Italy 
             \and 
	     Dipartimento di Fisica e Astronomia ``Galileo Galilei'', Universit\`{a} di Padova, Vicolo dell'Osservatorio 3, I-35122 Padova, Italy 
            \and 
             INAF - Osservatorio Astronomico di Brera, Via E. Bianchi 46, 23807 Merate (LC), Italy
	     \and 
	     NASA Sagan Fellow
	     \and
	     CHEOPS Fellow, SNSF NCCR-PlanetS 
}

   \date{Received September 15, 1996; accepted March 16, 1997}

 
  \abstract
   {
    Understanding stellar activity in solar-type stars is crucial for the physics of stellar atmospheres as well as for ongoing  exoplanet programmes.
   }
   {
    We aim to test how well we understand stellar activity using our own star, the Sun, as a test case. }
   {
   We perform a detailed study of the main optical activity indicators (Ca~{\sc ii}  H \& K, Balmer lines, Na~{\sc i} D$_{\rm 1}$ D$_{\rm 2}$,
   and He~{\sc i} D$_{\rm 3}$) measured for the Sun using the data provided by the HARPS-N solar-telescope feed at the Telescopio Nazionale Galileo.
   We make use of periodogram analyses to  study solar rotation, 
   and we use
   the pool variance technique to study the temporal evolution of active regions. 
   The correlations between the different activity indicators as well as the correlations between activity indexes and the derived
   parameters from the cross-correlation technique are analysed. We also study the temporal evolution of these correlations and their
   possible relationship with indicators of inhomogeneities in the solar photosphere like sunspot number  or radio flux values. 
 }
   {
   The value of the solar rotation period is found in all the activity indicators, with the only exception being H$\delta$. The derived
   values vary from 26.29 days (H$\gamma$ line) to 31.23 days (He~{\sc i}).  From an analysis of sliding periodograms
   we find that in most of the activity indicators 
   the spectral power is split into 
   several ``bands'' of periods around 26 and 30 days, that might be explained by the migration of active regions 
   between the equator and a latitude of $\sim$  30$^{\circ}$, spot evolution or a combination of both effects. 
   In agreement with previous works a typical lifetime of active regions of $\sim$ ten rotation periods is inferred from the pooled variance diagrams.
   We find that H$\alpha$, H$\beta$, H$\gamma$, H$\epsilon$, and He~{\sc i}  show a significant correlation with the $S$ index.
   Significant correlations between the contrast, bisector span, and the
   heliocentric radial velocity with the activity indexes are also found. 
   We show that the full width at half maximum, the bisector, and the disc-integrated magnetic field correlate with the radial velocity 
   variations.
   The correlation of the $S$ index and H$\alpha$  changes with time, increasing with larger sun spot numbers and solar irradiance.
   A similar tendency with the $S$ index - radial velocity correlation is also present in the data.
}
   {
   Our results are consistent with a scenario in which higher activity favours the correlation between the $S$ index and 
   the H$\alpha$ activity indicators and between the $S$ index and radial velocity variations.
}


   \maketitle
%
\section{Introduction}

 Our own star, the Sun, constitutes a benchmark in the study of stellar magnetic
 activity. Unlike other stars, the solar surface can be resolved, and direct information about
 the size, contrast, or the location of surface inhomogeneities can be obtained.

 Understanding stellar activity is crucial for the detection
 of small, rocky, potentially habitable planets around low-mass stars
 \citep[e.g.][]{2016PASP..128f6001F}.
 The detection of Earth-twins via the radial velocity technique requires
 a radial velocity precision of the order of 10 cms$^{\rm -1}$ which is an order of magnitude lower
 than the radial velocity variations induced by the presence of inhomogeneities
 (cool spots, hot faculae, plages) in the stellar surface, typically  
 between 1 and 200 ms$^{\rm -1}$ with timescales of 2-50 days in solar-type stars
 \citep[e.g.][]{2016ASSL..428....3H}. 
 Regarding the Sun, its
 apparent radial velocity variations as a star are between 1 and 20 ms$^{\rm -1}$ over timescales from days to years
 \citep{2010A&A...519A..66M,2015ApJ...814L..21D,2016MNRAS.457.3637H,2016A&A...587A.103L}.
 Variations as large as 200 ms$^{\rm -1}$ are localised into
 spotted and facular regions \citep{2010A&A...519A..66M}, 
 but they are much reduced when averaged over the whole solar disc.

 With the aim of a better understanding of stellar signals
 the small solar telescope at the Telescopio Nazionale Galileo (TNG)  
 is able to obtain precise full-disc radial velocity measurements of the
 Sun using the HARPS-N spectrograph 
 \citep{2015ApJ...814L..21D,2016SPIE.9912E..6ZP,2019ApJ...874..107M,2019MNRAS.tmp.1180C}.
 The approach is to observe the Sun as a star, allowing us to directly
 correlate any change in the observed surface inhomogeneities 
 with variations in the full-disc radial velocity.

 The HARPS-N solar data offer an unique opportunity to achieve a deeper
 understanding of how solar (and by extension stellar) activity produces apparent radial velocity variations. 
 Therefore, in this paper we present
 a detailed analysis of the main optical activity indicators
 (Ca~{\sc ii} H \& K,
   Balmer lines, Na~{\sc i} D$_{\rm 1}$ D$_{\rm 2}$,
     and He~{\sc i} D$_{\rm 3}$)
 measured during the first three years of the solar telescope operations.

 This paper is organised as follows. Section~\ref{observations_sect} describes
 the observations. The periodogram analysis and the correlations between different
 activity indicators is presented in Sect.~\ref{data_analysis}.
 The results are discussed at length in Sect.~\ref{disc_section}.
 Our conclusions follow in Sect.~\ref{summary}.

\section{Observations} 
\label{observations_sect}


\begin{table}
\centering
\caption{
Bandpasses for the activity indexes considered in this work.}
\label{table_act_index}
\begin{tabular}{lrr}
\hline
\bf{S index}$^{\rm (a)}$  & Wavelength ($\AA$) & Width($\AA$) \\
\hline
Red wing    & 3901.07 & 20   \\
K emission  & 3933.67 & 3.28 \\
H emission  & 3968.47 & 3.28 \\
Blue wing   & 4001.07 & 20   \\ 
\hline
\bf{H$\alpha$}$^{\rm (b)}$  & Wavelength ($\AA$) & Width($\AA$) \\
\hline
Red wing            & 6580.31 & 8.75 \\
H$\alpha$ emission  & 6562.81 & 1.60 \\
Blue wing           & 6550.87 & 10.75 \\
\hline
\bf{H$\beta$}$^{\rm (c)}$  & Wavelength ($\AA$) & Width($\AA$) \\
\hline
Red wing            & 4845.0  &   10 \\
H$\beta$ emission   & 4861.32  & 1.60 \\
Blue wing           & 4880.0   &  10 \\
\hline
\bf{H$\gamma$}$^{\rm (c)}$  & Wavelength ($\AA$) & Width($\AA$) \\
\hline
Red wing            &  4320. &   10 \\
H$\gamma$ emission   & 4340.46 & 1.60 \\
Blue wing           &  4360.  &  10 \\
\hline
\bf{H$\delta$}$^{\rm (c)}$  & Wavelength ($\AA$) & Width($\AA$) \\
\hline
Red wing            &  4085. &   10 \\
H$\delta$ emission   & 4101.76 & 1.60 \\
Blue wing           &  4120.  &  10 \\
\hline
\bf{H$\epsilon$}$^{\rm (c)}$  & Wavelength ($\AA$) & Width($\AA$) \\
\hline
Red wing             &  3901.07 &   20 \\
H$\epsilon$ emission & 3970.07 & 1.60 \\
Blue wing            &  4001.07  &  20 \\
\hline
\bf{ He~{\sc i} D$_{\rm 3}$ } $^{\rm (b)}$  & Wavelength ($\AA$) & Width($\AA$) \\
\hline
Red wing             &  5881.0 &  5 \\
He emission           & 5875.62 & 0.4 \\
Blue wing            &  5869.0  &  5 \\
\hline
\bf{ Na~{\sc i} D$_{\rm 1}$ D$_{\rm 2}$ }$^{\rm (b)}$  & Wavelength ($\AA$) & Width($\AA$) \\
\hline
Red wing             &  5805.0 &  20 \\
D$_{\rm 1}$ emission & 5895.92 & 0.5 \\
D$_{\rm 2}$ emission & 5889.95 & 0.5 \\
Blue wing            &  6090.0  &  20 \\
\hline
\end{tabular}
\tablefoot{$^{\rm (a)}$ \cite{1996AJ....111..439H}; $^{\rm (b)}$ \cite{2011A&A...534A..30G};
$^{\rm (c)}$ \cite{2008PASP..120.1161W}}
\end{table}

 To date, the Sun has been observed for a period of 1049 days (i.e., $\sim$ 2.9 yrs) from 
 BJD = 2457218 (July 14, 2015) to BJD = 2458267 (May 29, 2018), corresponding to the
 late phase of the 24th solar cycle.
 A total of 48335 HARPS-N observations were collected during this period.
 The median number of observations per day is 45, although in some days
 the number of observations can be as high as 80.

  HARPS-N spectra cover the
  wavelength range 383-693 nm with a resolving power of $R$ $\sim$ 115000.
  Data were reduced using the latest version of the Data Reduction Software
  \citep[DRS V3.7,][]{2007A&A...468.1115L} which
  implements the typical corrections
  involved in \'echelle spectra reduction, i.e. bias level, flat-fielding,
  order extraction, wavelength calibration, and merge of individual orders. 
  Radial velocities (RVs) are computed by
  cross-correlating the spectra of the target star with an optimised
  binary mask \citep{1996A&AS..119..373B,2002A&A...388..632P}.
  For the Sun the G2 mask was used.
  The HARPS-N DRS also provides
  several CCF asymmetry diagnostics,
  such as the CCF width (FWHM), the bisector span (BIS), and the CCF contrast. 
  Typical values of the signal-to-noise ratio (SNR) around 5500 \AA \space are
  $\sim$ 380, although in the period between BJD 2458060 and 2458160 
  the achieved SNR decreases due to a damaged optical fiber and typical values are around 130. 
 
  Activity indexes in the main optical indicators i.e., Ca~{\sc ii} H \& K,
  Balmer lines (from H$\alpha$ to H$\epsilon$), Na~{\sc i} D$_{\rm 1}$ D$_{\rm 2}$,
  and He~{\sc i} D$_{\rm 3}$ were computed.
  Our definition of the bandpasses for the Ca~{\sc ii} H \& K
  activity $S$ index is made following \cite{1996AJ....111..439H}, using a triangular
  filter in the core of the lines.  We note that this is the only index for which
  a triangular shape is used.
  $S$ index values were corrected for the presence of ghosts and transformed to the
  Mount Wilson scale  using the relationship provided by \cite{2011arXiv1107.5325L}.
  We note
  that the use of R'$_{\rm HK}$ is not needed for the purpose of this work.
  Our scope is to compare different activity indicators and analogous quantities to the
  R'$_{\rm HK}$ have not been defined for other activity-sensitive lines.
  For the H$\alpha$, He~{\sc i}, and Na~{\sc i}  indexes the definitions by  \cite{2011A&A...534A..30G}
  were followed. For the rest of the Balmer lines
  the prescriptions given by \cite{2008PASP..120.1161W} were used.
  In some cases, small changes in the width of the continuum passbands
  were introduced. Our bandpasses are summarised in Table~\ref{table_act_index}.
  Fluxes were measured using the IRAF\footnote{IRAF
  is distributed by the National Optical Astronomy Observatories,
  which are operated by the Association of Universities for
  Research in Astronomy, Inc., under cooperative agreement with
  the National Science Foundation.} task {\sc sbands}.  
  Before
  measuring the fluxes, each individual spectrum was corrected for its
  corresponding radial velocity
  (see below)
  using the IRAF task {\sc dopcor}.
  Uncertainties
  in the activity indexes
  are computed from the uncertainties in each bandpass which 
  are computed as:

  \begin{equation}
  \Delta F = \frac{1}{\sqrt{n}} \left[\frac{F}{SNR}\right] 
  \end{equation}

 \noindent where $F$ is the mean flux in the band, SNR is the signal-to-noise ratio, and
 $n$ is the number of integration points. Here, we assume that both the flux and the SNR are constants
 in each bandpass, which is a reasonable assumption given their small width. 
 A conservative approach was taken and for each index the SNR of the red bandpass
 was chosen as the representative value for all the bandpass. 
 Nevertheless, given the high SNR of our spectra, uncertainties in the activity
 indexes are quite small, of the order of  0.16\% for the $S$ index and $\sim$ 0.04\% for
 the H$\alpha$ index.

  The solar RVs derived by the DRS 
  are 
 dominated by the motion of the Solar System's giant
  planets (Jupiter and Saturn). 
  The RV effects of these motions are removed 
following the procedure
 by \cite{2019MNRAS.tmp.1180C}
  which can be consulted for further details.
  
\begin{figure}[!htb]
\includegraphics[scale=0.45]{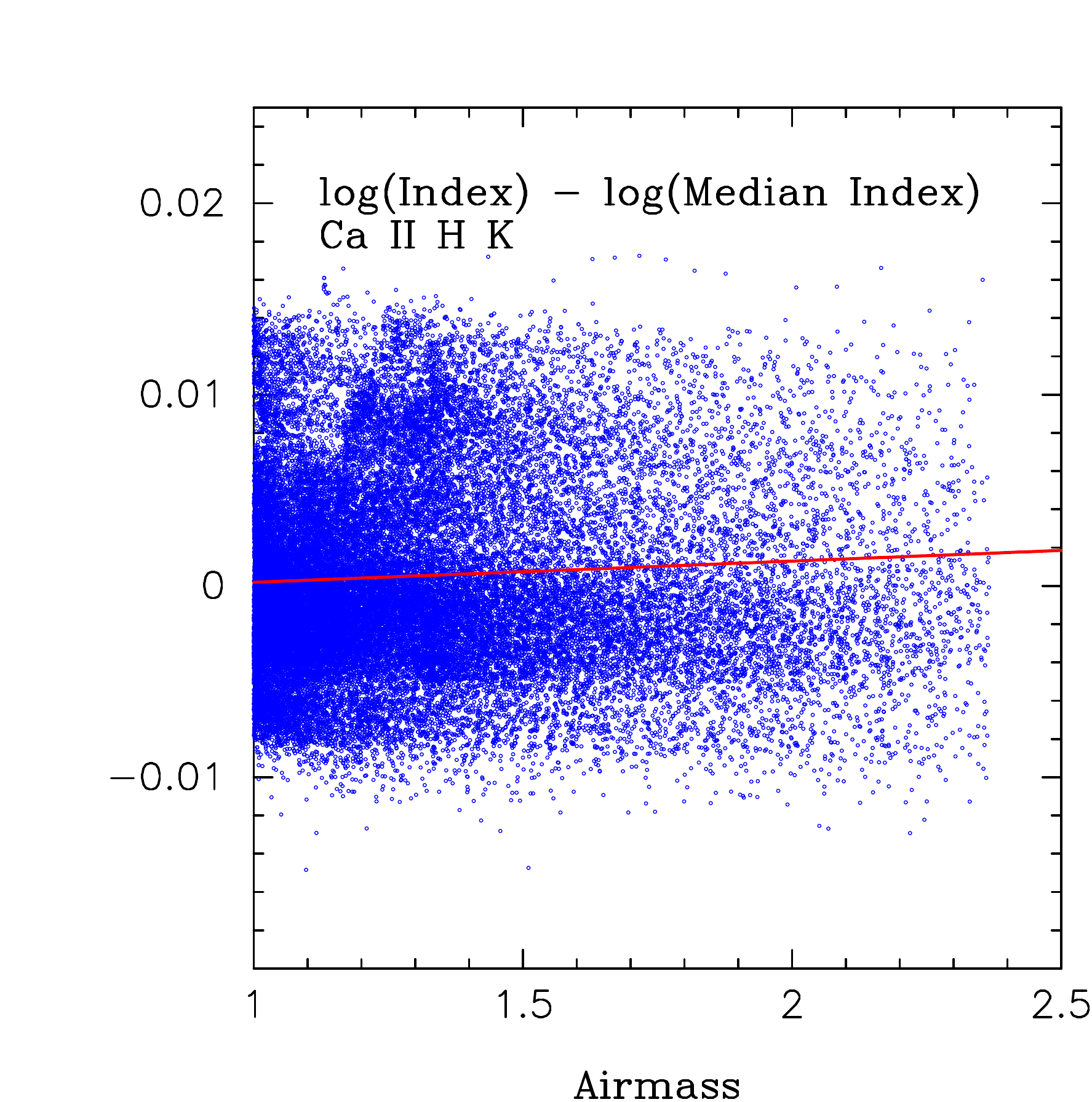}
\caption{
Airmass effect correction for the Ca~{\sc ii} H \& K index.
}
\label{airmass_corr}
\end{figure}

  In order to remove outliers in the different activity indexes a 3$\sigma$ clipping procedure was applied for
  all indexes. A quick inspection of the activity indexes reveals a daily variation correlated with the
  airmass of the observations.
  In order to correct for this effect we compute a daily median index.
  The quantity ($\log (index)$ - $\log (median daily index)$) is then plotted against
  the airmass and a linear fit is performed, see Figure~\ref{airmass_corr} for the Ca~{\sc ii} H \& K index.
  We note that the daily median values are only used to keep the indexes in their original scale 
  values thus favouring possible comparisons with other works or stars. The fit is used to correct 
  for the airmass effect. Finally, the median values are added again.
   It is clear from the figure that the airmass effect has a minor impact on the indexes. Nevertheless, we prefer to
  perform this correction in order to avoid possible annual periods in the analysis.
  We note that the scatter in the $S$ index data is $\sim$ 0.0019 (value corresponding to the standard deviation of the data)
  which is consistent with a scatter dominated by photon noise error. 
   
\section{Data analysis}\label{data_analysis}
\subsection{Periodogram analysis}

\begin{figure*}[!htb]
\centering
\begin{minipage}{0.33\linewidth}
\includegraphics[scale=0.40]{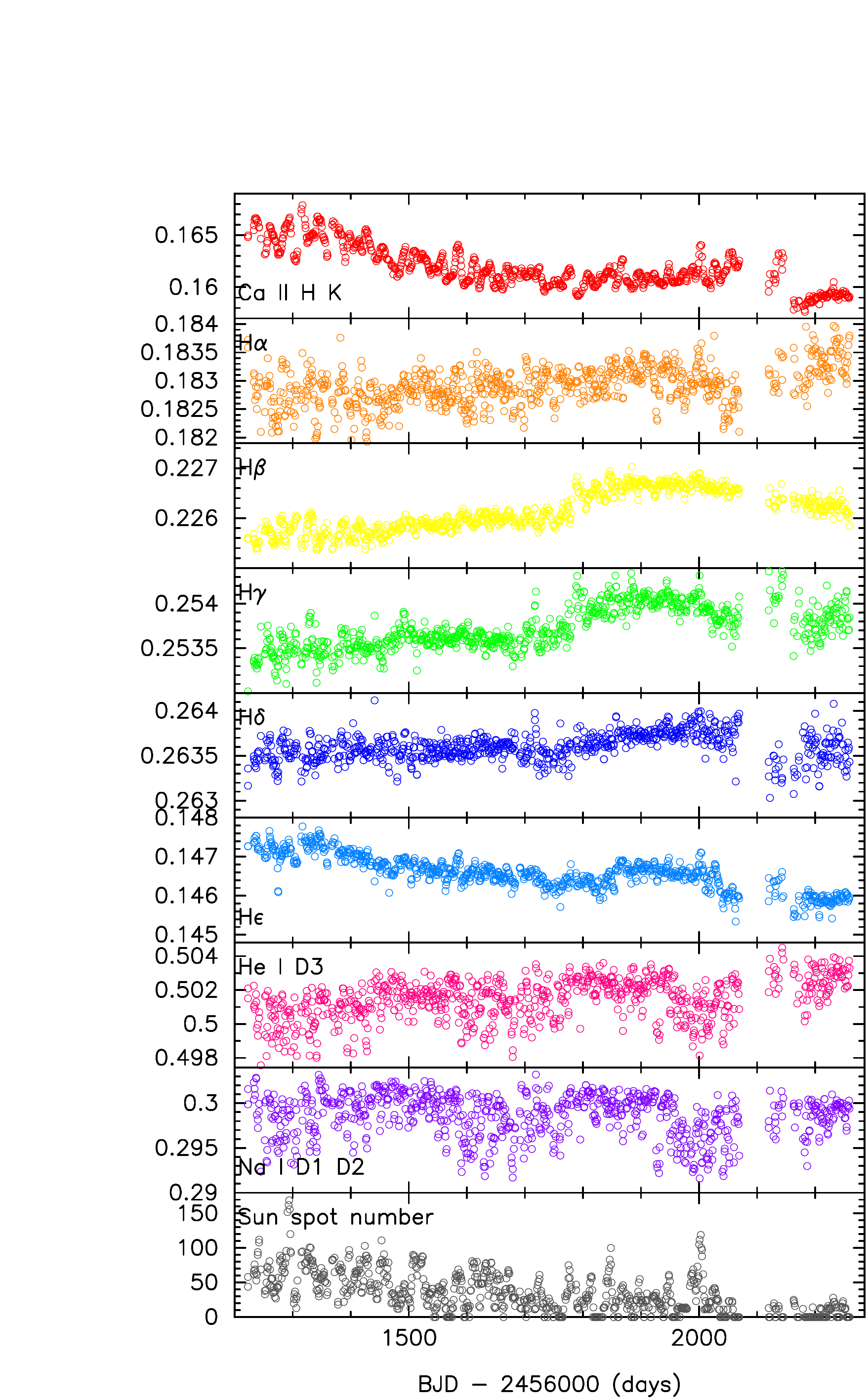}
\end{minipage}
\begin{minipage}{0.33\linewidth}
\includegraphics[scale=0.40]{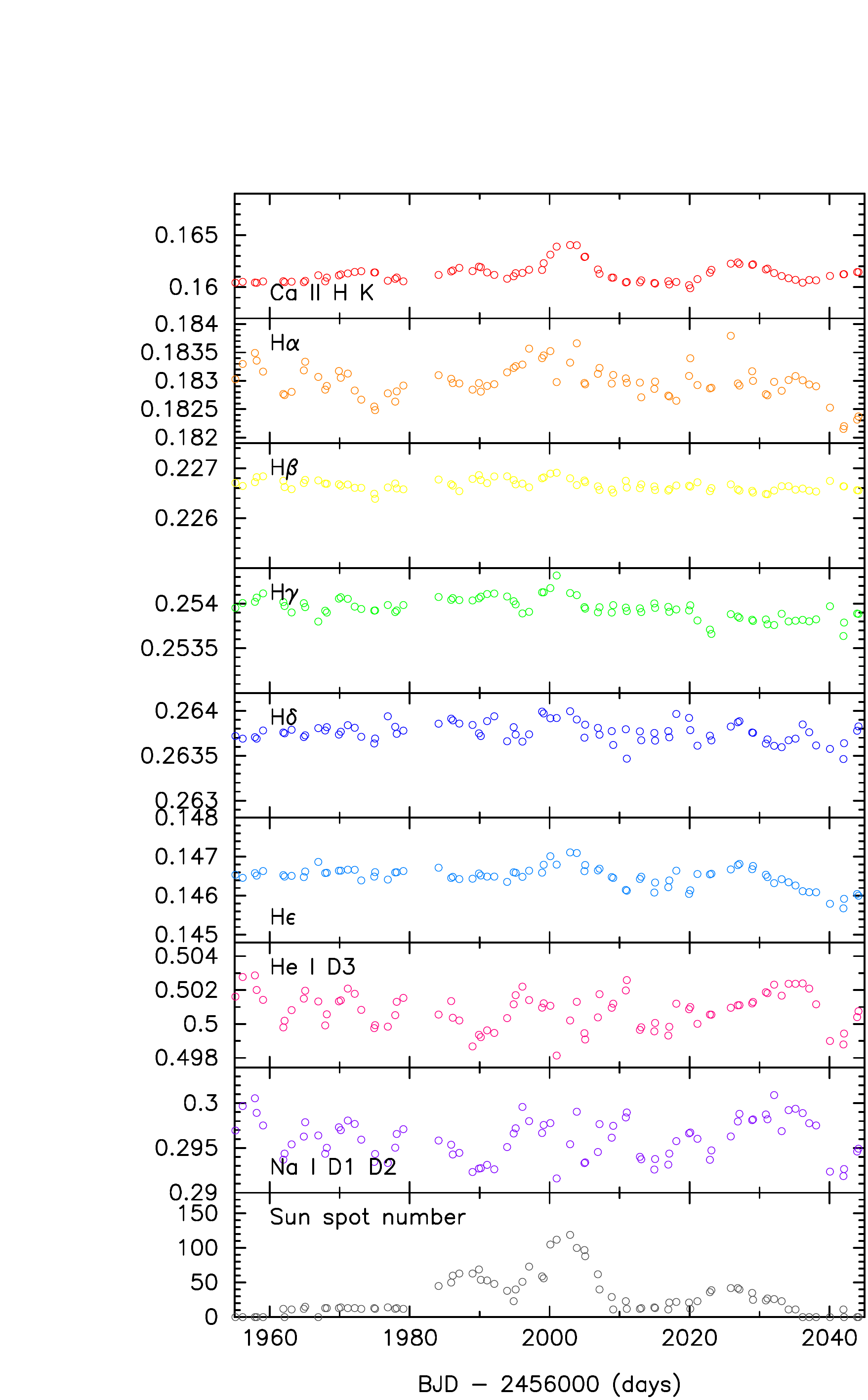}
\end{minipage}
\begin{minipage}{0.33\linewidth}
\includegraphics[scale=0.40]{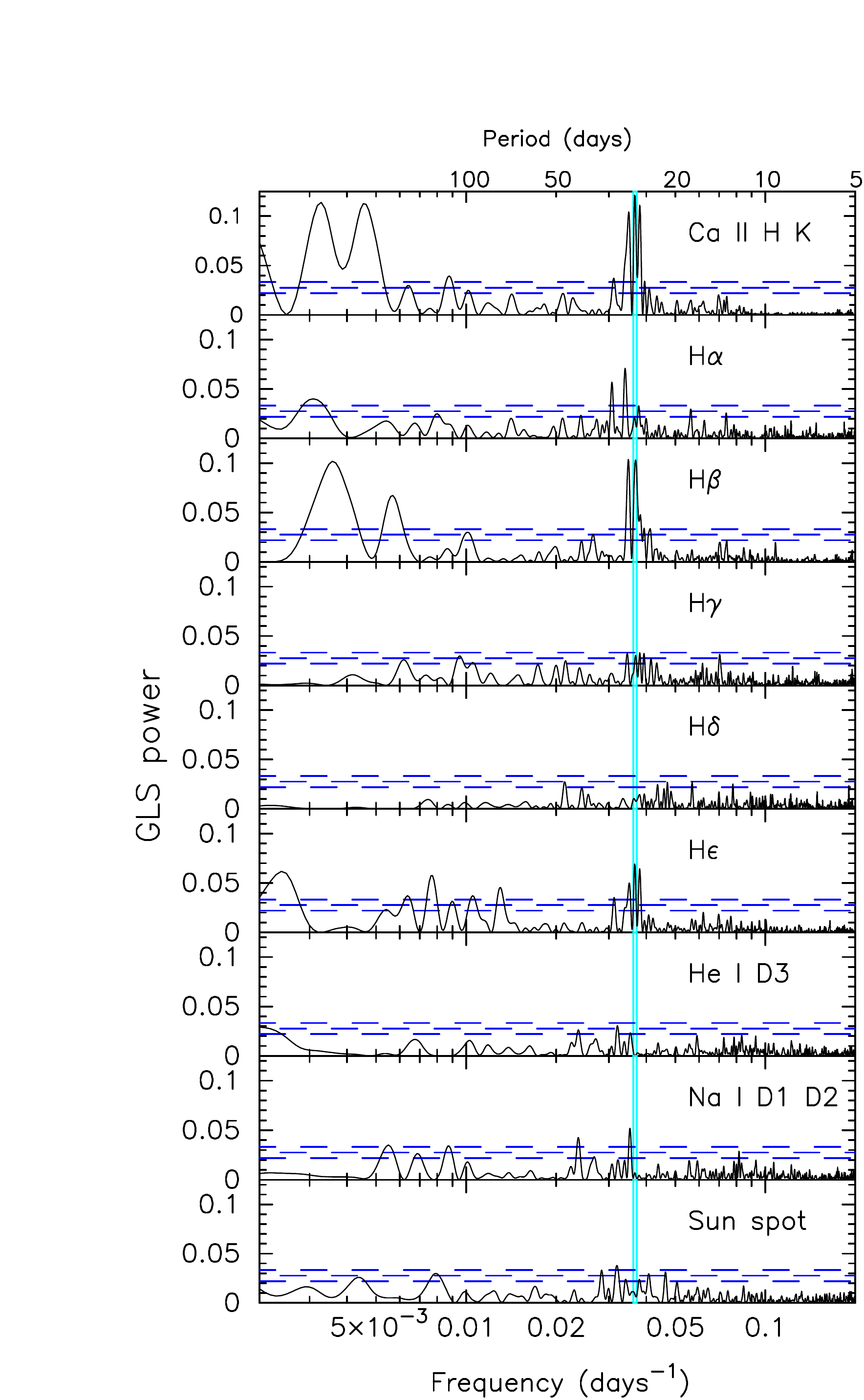}
\end{minipage}
\caption{  
Left: Activity indexes (defined in the text) as a function of time. Daily median values are shown.
Centre: Zoom on the time series covering a period of two rotations.
Right: Corresponding generalised Lomb-Scargle periodograms (after the subtraction of the main long-term
periods, see text for details). The light blue lines indicate the solar synodic
period of 27.2753 days \citep{1977asqu.book.....A} while values corresponding to a FAP
of 10\%, 1\%, and 0.1\% are shown with dashed lines.}
\label{activity_indices}
\end{figure*}

%

  Figure~\ref{activity_indices} (left) shows the temporal variations of the analysed activity indexes.
  The median index per day is shown.  From now on, we refer to the median index per day in all the
  analysis performed.
  A search for periodicities 
  was performed by using
  the generalised Lomb-Scargle periodograms \citep[GLS,][]{2009A&A...496..577Z}.
  The periodograms are dominated by long-term signals. 
  In order to
  clearly identify signals at the solar rotation period, we subtracted the main long-period trends for each index
 by fitting  a sinusoidal function (a procedure
  usually referred to as prewhitening).
  The subtracted periods are $\sim$  27000 days for Ca~{\sc ii}, and H$\epsilon$, 
  $\sim$ 3800 days for H$\alpha$,
  and of the order of $\sim$ 1200 days
  for H$\beta$, H$\gamma$ and H$\delta$.
  Additional long-term signals with periods between 200 and 600 days were removed from  most of the indexes.
   The subtracted periods as well as the number of prewhitenings performed are given in Table~\ref{table_sli_analysis}
  while
  Figure~\ref{activity_indices} (right) shows the corresponding GLS periodograms.
  Periods were subtracted in a sequential way, until a period inside the time interval of the Sun's rotation
  (between 25-34 days) dominates the periodogram. For example, for the S index,
   the period labeled as P$_{\rm 1}$ in Table ~\ref{table_sli_analysis} was subtracted,
  while for H$\beta$ 
  the periods labeled as P$_{\rm 1}$ and P$_{\rm 2}$ were prewhitened.
  We note that the time series are those of the original datasets, i.e, before any prewhitening procedure is applied,
  while we show the periodograms after the described prewhitening procedure.
  Several conclusions can be drawn from this figure. First, long-periods still remain, in particular
  for the H$\beta$ index. 
  Second, the solar rotation is found in all activity indicators with the exception
  of H$\delta$ 
  where it is not clearly detected 
  and which shows a lot of periods close to 
   $\sim$ 22 days. 
  The most significant periods that we identify with the solar rotation for each index
  are given in Table~\ref{table_sli_analysis}. Their uncertainties are those provided by the
  GLS analysis. We note that these are synodic periods and therefore not corrected from effects
  due to the  Earth's revolution around the Sun. 
  Our results show that the rotation period depends on the indicator used. It varies from
  26.29 days (H$\gamma$ index) to 31.23 days (He~{\sc i}).

  Figure~\ref{activity_indices} also reveals an increase in the H$\beta$, H$\gamma$, and to a lesser extent H$\delta$
  values at BJD $\sim$ 2457800 days. Then, around BJD 2458100 days the indexes show a small decline.  
  These values do not correspond to the period of reduced SNR.
   We checked whether large plages were present in the solar surface at these epochs 
  by carefully analysing the USET images\footnote{http://sidc.oma.be/uset/searchForm.php}
  at the Solar Influence Data Analysis Center of the Royal Observatory of Belgium
  as well as the NASA/SDO AIA 1600 images\footnote{https://sdo.gsfc.nasa.gov/data/aiahmi/}.
  We find that while at BJD $\sim$ 2457800 days the solar surface does not differ in a significant way from other epochs,
  at BJD $\sim$ 2458000 days (where the H$\beta$ index shows a peak)
  several active regions (with associated NOAA numbers from 12673 to 12677) formed by plages and dark
  spots are visible across the solar surface. 
  
  Furthermore, for the He~{\sc i}, Na~{\sc i}, and perhaps H$\alpha$ lines several periods of low values can be seen.
  As they are separated by $\sim$ 350 days we suspect that this effect could be related to the  presence of weak telluric lines
  contaminating the cores or continua of these indexes and thus inducing annual variations.

\begin{figure*}[!htb]
\centering
\begin{minipage}{0.49\linewidth}
\includegraphics[scale=0.45]{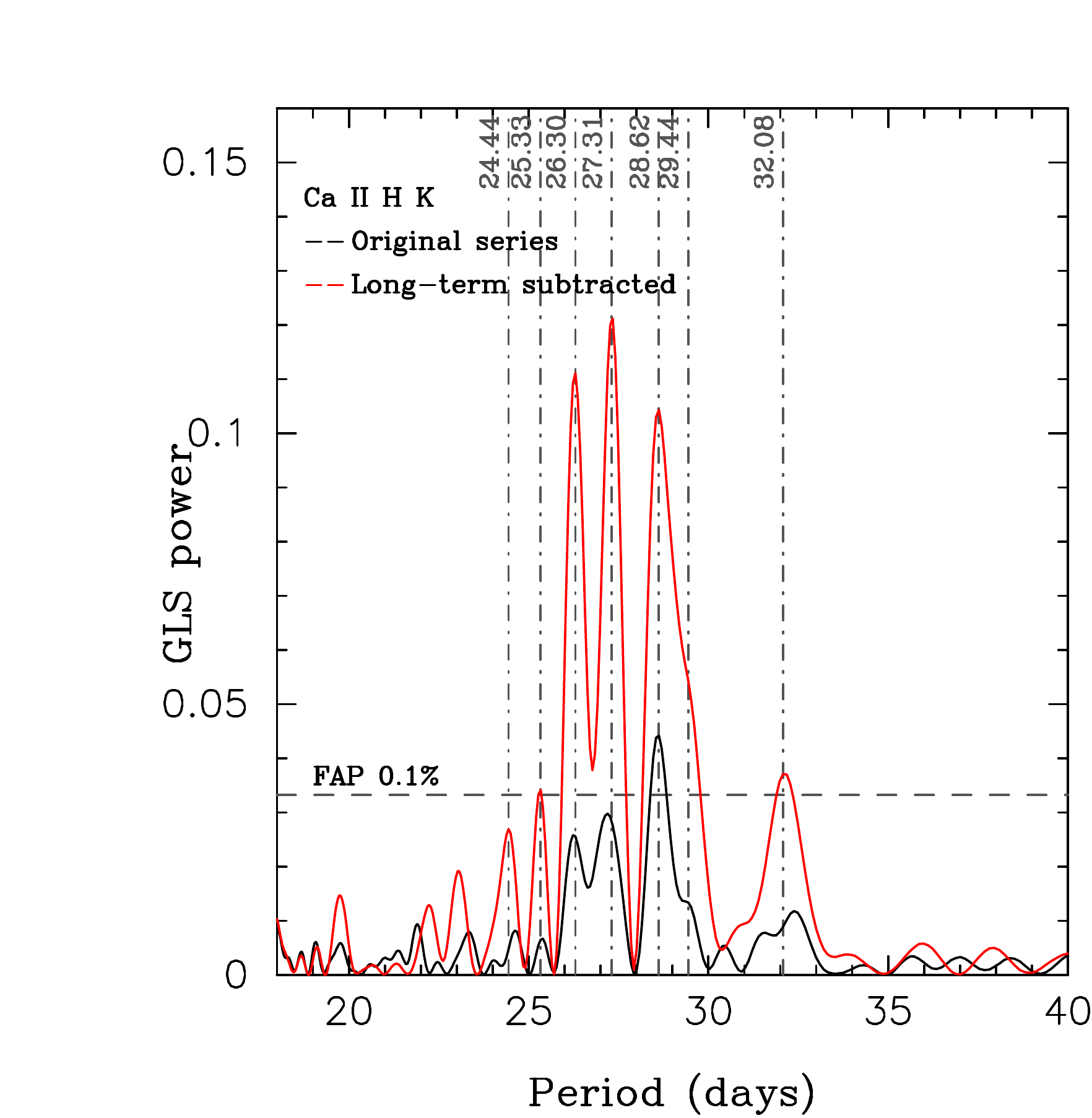}
\end{minipage}
\begin{minipage}{0.49\linewidth}
\includegraphics[scale=0.45]{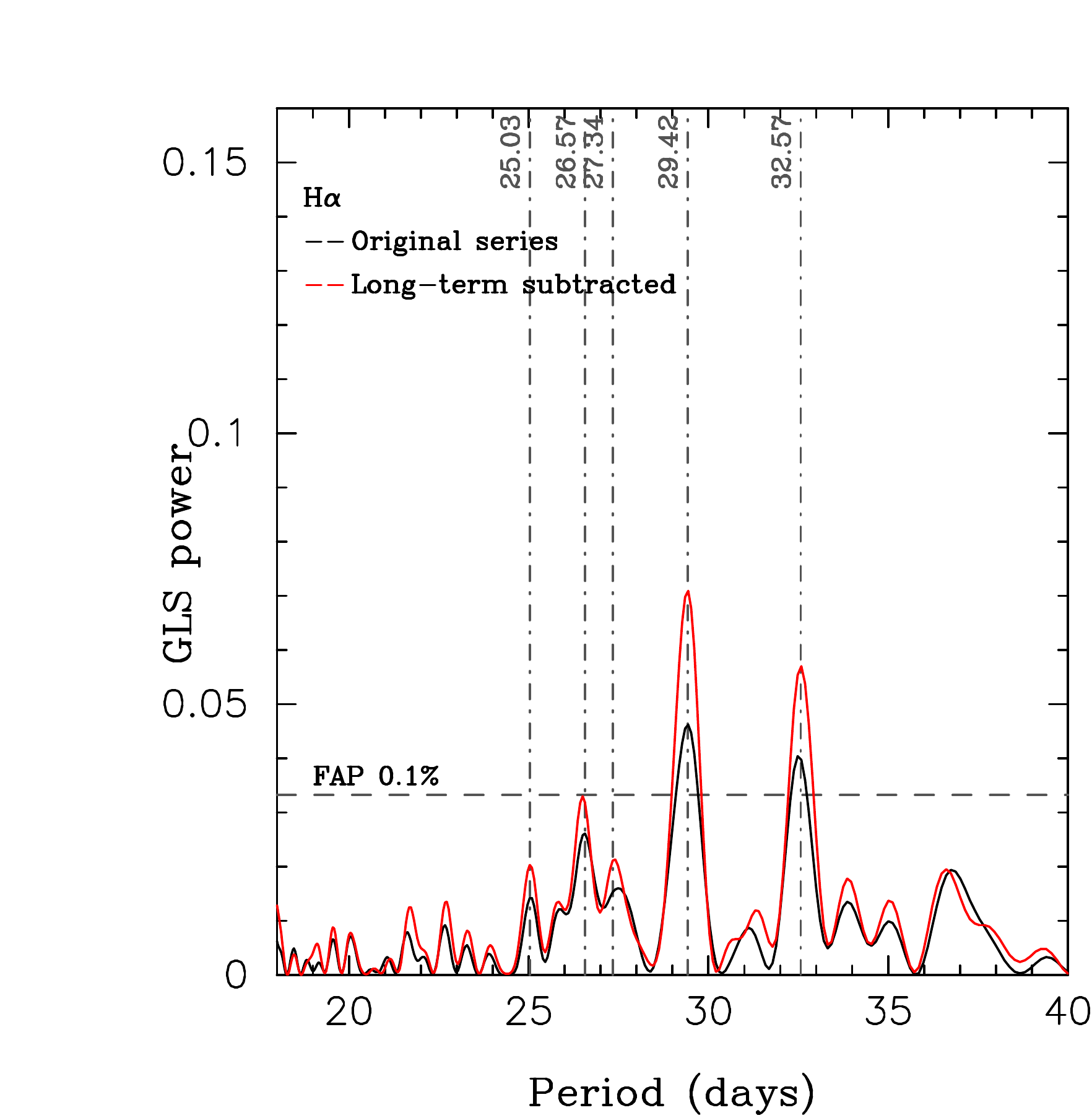}
\end{minipage}
\caption{   
GLS periodograms of the S (left) and H$\alpha$ indexes (right) after each prewhitening zoomed around the periods of interest.
Vertical dashed-to-dot lines show the periods discussed in the text.
}
\label{caiihk_periodogram_zoom}
\end{figure*}

  We finally study the impact of the prewhitening in the determination of the rotation period.
  This is illustrated in Fig.~\ref{caiihk_periodogram_zoom} where we show the periodograms corresponding
  to the S index (left panel) and H$\alpha$ index (right panel) after each prewhitening. The plot shows a zoom around the region of the solar
  rotation. 
  It is clear from the figure that the prewhitening procedure does not change the found periods,
  only their relative strength. 
  The GLS power of the periods increase after each prewhitening. 
  The periodogram corresponding to the S index show a more complex pattern than in the H$\alpha$ case.
  For the S index three main periods at $\sim$ 26.30, 27.31 and 28.62 days are clearly dominant in all cases.
  The latest period is broad and seems to be mixed with a close period at 29.44 days. 
  In addition, two secondary peaks appear at $\sim$ 25.33 and 32.08 days, respectively.
  Regarding H$\alpha$, another three period structure can be seen that dominates the periodogram with power at $\sim$ 26.57, 29.42, and 32.58 days. 
  
  We note that the period at 32 days which is dominant in H$\alpha$ is also seen in the S index, although it is not among the 
  dominant signals for this indicator.
  The other main difference between the S index and H$\alpha$ periodograms is that the periods at $\sim$ 27.3 days and $\sim$ 26.3 days 
  which are dominant in the S index are much less significant in the H$\alpha$ data.

\subsection{Temporal evolution of active regions}
\label{stacked_periodograms}

 In order to study the temporal evolution of active regions we made use
 of the pooled variance technique \citep{1997SoPh..171..191D,1997SoPh..171..211D,2004A&A...425..707L,2017A&A...598A..28S}.
 In brief, for a given time interval, $t$,  the data are binned into segments of length 
 $t$, and the average variance of the set of these segments is computed.
 It is expected that as the time scale increases, the effects from processes
 at longer time scales become more noticeable and therefore the pooled variance increases.
 
 For example, the light curve of a star with a fixed, non evolving, spot  pattern and rigid rotation 
  will be perfectly periodic. The variance of this light curve will
 increase with $t$, until $P_{\rm rot}$ is reached. For timescales longer
 than the rotation period, the mean variance will stay approximately constant.
 If we now introduce the evolution of active regions, and the timescale
 of evolution is significantly longer than $P_{\rm rot}$, 
 the variance of the light curve will increase for timescales longer than 
 $P_{\rm rot}$ until the mean timescale of evolution of the active regions is reached.
 We will have a change in slope because the increase of the variance with $t$ is generally different from that produced by rotation,
 given that the shape of the rotational modulation is different from the light changes associated with active region evolution.

\begin{figure*}[!htb]
\centering
\includegraphics[scale=0.50]{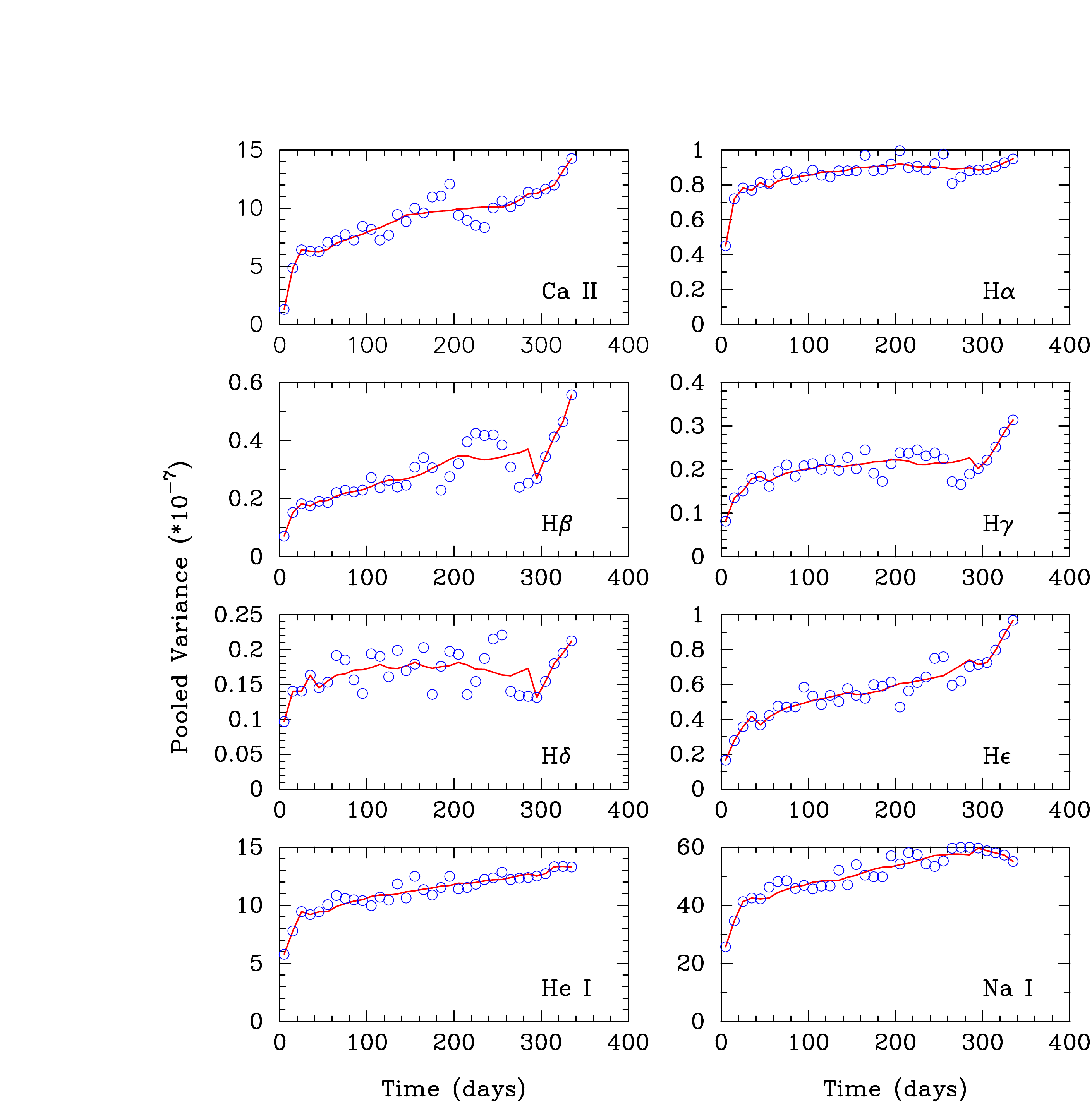}  
\caption{Pooled variance diagrams for the different activity indicators analysed in this work (\emph{see text}).
The red line is a smoothed function for ease reading of the plots.}
\label{pooled_variance}
\end{figure*}

 The corresponding pooled variance diagrams for the activity indexes analysed in this work
 are shown in Figure~\ref{pooled_variance}. The pooled variance was computed from
 $t$ = 5 days up to one third of the total time of the observations, with a time step of 10 days.
 The analysis of the data shows two slope changes, one around 30 days which corresponds to
 the solar rotation period and a second one around 300 days from which we infer a typical lifetime
 for active regions of $\sim$ 10 rotation periods. We note that the second slope change around 300 days
 is not so strong in the He~{\sc i} and Na~{\sc i} activity indicators. 
 The time scale of $\sim$ 300 days is quite similar to the mean life time of  complexes of activity dominated
 by faculae and plage 
 \citep[e.g.][]{1986SoPh..105..237C} 
 while  
 sun spots are known to live for hours to months \citep[e.g.][]{2003A&ARv..11..153S}.
  We finally note the peculiar behaviour of the H$\alpha$ index, which shows almost no variation
  after 300 days in contrast to the other indexes.


 We study  solar rotation via time-varying periodogram
 power of the activity indexes. 
 Power spectra are computed for each activity index time series in overlapping segments of 600 days,
 with a two day time shift between consecutive segments.
 Before computation, the long-term periods identified in the GLS analysis described
 above  were prewhitened in a sequential way.
 Additionally, we perform a linear detrending
 of the data in each 600 days temporal window independently. 
 The corresponding sliding periodograms
 are shown in Figure~\ref{peridogramas_hbeta_900}.

 We note that a large window is needed if we want to average the effect of
 the different surface inhomogeneities 
 \citep[e.g.][]{2012ApJ...761...11B}.
 Figure~\ref{peridogramas_window} shows the corresponding sliding periodogram for the S index
 computed using a temporal window of 300, 600, and 900 days, respectively.
 Although similar in form, i.e, the Scargle power is found at roughly the same periods, 
 it is clear from the figure than when using
 a small temporal window the revealed signals tend to be broader. 
 In addition, discontinuities appear for periods larger than $\sim$ 40 days.
 If a window of 900 days  (i.e., almost the full temporal coverage of our data)
 is considered (bottom panel) the periods appear significantly narrower, but many
 white vertical strips (meaning that no temporal window is centred at the corresponding date)
 appear in the plot.
  Note that gaps in the data should not been quite different from one window size to another.
 After all, all windows are quite large. However, with increasing the temporal window  we are covering a shorter
 global time scale (i.e, we are making a kind of zoom, see the range of dates in the X axe for the different plots
 of Figure~\ref{peridogramas_window}) so the gaps in the data are more easily
 seen in the plots.
 We choose 600 days as a compromise between having a large temporal window
 (so periods are narrow enough) and having a large temporal coverage.  
 We also investigated whether the prewhitening procedure performed to the data does have an
 impact on the sliding periodograms.  This is illustrated in Figure~\ref{peridogramas_prewhitening}
 where sliding periodograms for the S index are shown for the original time series (up panel), and after
 prewhitening the long-term signal (middle panel). 
 see Table~\ref{table_sli_analysis} for the exact values of the subtracted periods.
 We can see that the periodograms are nearly identical, with only very small differences. 
 We conclude that the prewhitening procedure
 has a small impact on the periodograms.

 We use the periodograms to search for  
 a signature of differential rotation. 
  Table~\ref{table_sli_analysis} shows the spread in rotation periods,
 $\alpha$, defined as
 $(P_{Max} - P_{Min})/{P_{Max}}$ being $P_{Max}$ and $P_{Min}$ the maximum
 and minimum rotation periods.
 It was measured by visual inspection of the 
 drift in the mean frequency of the forest of peaks around the rotation frequency.
  Uncertainties on $\alpha$ are computed by assuming an error on $P_{Max}$ and $P_{Min}$
 of 0.07 days (median values of the errors in the periods listed in Table~\ref{table_sli_analysis})
 and applying uncertainties propagation.
  These uncertainties are clearly underestimated. 
 This is because our real uncertainty depends on the choice of $P_{Max}$ and $P_{Min}$ which in turn
 rely on the noise level of each index.
 This effect might lead to a lower $\alpha$ value as peaks on the outside are in general weaker and
 most likely to be missed. Furthermore, a period which is present in one index might not be 
 present in another index with a higher level of noise. 
 The value of $\alpha$ depends significantly on 
 the index. It is higher for the $S$ index  and H$\epsilon$.
 We note that for H$\delta$ and He~{\sc i} we were not
 able to determine $\alpha$. 
 That (in addition to the previous results) suggests that either the H$\delta$ line is less sensitive to 
 stellar activity than other indicators or that the definition of this index should
 be revised. Finally, for H$\gamma$ the value of $\alpha$ 
 was derived from the first harmonic of the rotation period.

  It is important to note that we cannot unambiguously
  attribute the splitting of the rotation period peaks to differential rotation.
  It is well known that spot or active region evolution can induce similar beat patterns,  broadening or splitting peaks in the periodograms
  \citep[e.g.][]{1994A&A...290..861L,2015MNRAS.450.3211A}.
   As discussed, we measured typical  activity complex lifetimes of 10 rotation periods. However,  the presence
  of small-scale features evolving over shorter timescales cannot be excluded.
  The presence of several close periods might, indeed, be a combination of differential rotation and spot evolution.

  In addition to the periods inside the 20-36 days time interval of the Sun's rotation, other periodicities
  are found in the sliding periodograms,  see Figure~\ref{peridogramas_hbeta_900}.
  In particular, one period at $\sim$ 14.5 days is visible in most of the indicators, specially in H$\gamma$.
  When considering the $S$ index, a secondary peak at $\sim$ 13.5 days is also visible.
  We believe that these features might correspond to a 12-14 days periodicity previously identified in the literature.
  This period has been shown 
   to be an effect of an active longitude, i.e.,
  the result from two different
  active regions separated by 180 degrees in longitude \citep{1988AnGeo...6..417D,1990SoPh..130..369D,1992SoPh..142..365B}.
  The harmonic of the rotation period is also present and both effects contribute to this signal.
  It is also partly caused by the light curve of a single spot being similar to a truncated sinusoid
  (it is flat when the spot is on the invisible hemisphere).

  A period close to $\sim$ 18 days is seen 
  in H$\alpha$. To the best of our
  understanding, no reference to such a period
  has  been found in the literature.
  Periodicities around $\sim$ 43.5 and 47.5 days are visible in the $S$ index analysis, in agreement
  with previous works \citep[e.g.][]{2001A&A...377..297Z}. For H$\gamma$ a periodicity close to $\sim$ 51 days
  is found. The origin of this periodicity at 51 days is still unclear, although it seems that it appears most
  strongly in measurements related to emerging magnetic flux \citep{1992SoPh..142..365B}.

  Finally, several signals at larger periods are found. For the $S$ index, there is a signal ranging from 70 to 76 days,
  as well as another one from 86 to 92 days.
   Additional signals around $\sim$ 125, 135 days also appear in the periodogram of the S index.
  A clear signal at 100 days is seen in H$\gamma$, while H$\epsilon$ show
  a signal at $\sim$ 80 days and another one ranging from $\sim$ 98 to 113 days. 
  Signals at these periods have already been reported in other solar studies \citep[e.g.][]{2001A&A...377..297Z}.


  Another clear signal at $\sim$ 170 days appears in the periodogram of the S index.
This periodicity is a bit longer to the so-called Rieger cycles discovered in the periodicity of high-energy solar flares 
  \cite{1984Natur.312..623R} and it has been also found in indicators for solar activity such as sunspot areas \citep[e.g.][]{1998Natur.394..552O}.
  It has a periodicity around 155-160 days, and it usually appears only near the maxima
  of solar cycles. 
  \cite{2010ApJ...724L..95Z} suggested that this periodicity might be connected to the dynamics of magnetic
  Rosbby waves in the tachocline.

  We finally comment on the large period of 31.23 days derived from the analysis of the He~{\sc i} time series.
  This line is known to be formed at somewhat higher temperatures than other optical diagnostics of activity
  \citep{1997A&A...326..741S} and it is often used to study prominences and flares \citep{1988Sci...242.1586Z}.
  We note that such a period would correspond to active regions located at significant high latitudes, $\sim$ 60$^{\circ}$
  (but see next Section). 
  From Fig.~\ref{activity_indices} it can be seen that its significance level is lower 
  (false alarm probability, FAP $\sim$ 1\%)
  than the periods
  corresponding to other indexes that show FAP values highly below  0.1\%.
  This is also evident in the corresponding sliding periodogram, Fig.~\ref{peridogramas_hbeta_900} 
  (bottom left panel)
  where the
  period is not seen, as only periods with FAPs lower than 0.01 are shown.
  Instead, two periods are clearly visible, although only at the end of the observing window.
  These are the period at 12-14 days discussed above, and another period at $\sim$ 40 days.

\begin{table*}
\centering
\caption{
 Periods found in the GLS periodogram analysis, assumed
  rotation periods, and  relative spread in rotation period $\alpha$, potentially indicative of differential rotation, 
measured from the sliding
periodograms for the different activity indicators.
 Periods shown in italics are unlikely to be realistic. They should be understood as a way to represent and remove long term variability. 
} 
\label{table_sli_analysis}
\begin{tabular}{lrrrrrrcc}
\hline
 Index                              &  P$_{\rm 1}$  &    P$_{\rm 2}$    &    P$_{\rm 3}$ &    P$_{\rm 4}$  &     P$_{\rm 5}$ &   P$_{\rm 6}$              &   P$_{\rm rot}$         &   $\alpha$     \\
				    &  (days)       &    (days)         &      (days)    &     (days)      &        (days)   &    (days)                  &   (days)               & (\%)           \\
\hline                                                                  
 Ca~{\sc ii} H \& K                 & {\it 27100.06} &   27.32           &                &                 &                 &                            & 27.32 $\pm$ 0.06      & 4.97 $\pm$         0.25 \\
 H$\alpha$                          & {\it 3870.01}  &       29.42       &                &                 &                 &                            & 29.42 $\pm$ 0.09      & 2.44 $\pm$         0.24 \\
 H$\beta$                           & {\it 1182.67}  &   {\it   473.21}  &       28.69    &                 &                 &                            & 28.69 $\pm$ 0.07      &  2.63 $\pm$         0.25 \\
 H$\gamma$                          & {\it  1280.30} &   {\it   444.17}  &      288.01    &        75.11    &        26.29    &                            & 26.29 $\pm$ 0.08      & 0.79 $\pm$         0.26 \\
 H$\delta$                          & {\it 1029.74}  &   {\it 511.25}    &      310.59    &       199.35    &        42.95    &          46.84             &            -          &  -   \\
 H$\epsilon$                        & {\it 39958.61} &   {\it 606.49}    &      305.62    &        27.33    &                 &                            & 27.33 $\pm$ 0.06      & 4.64 $\pm$         0.25 \\
 He~{\sc i} D$_{\rm 3}$             &   363.09       &   {\it  6908.08}  &      114.99    &        31.23    &                 &                            & 31.23 $\pm$ 0.15      &  -   \\
 Na~{\sc i} D$_{\rm 1}$ D$_{\rm 2}$ &   358.79       &   {\it  7949.04}  &       28.36    &                 &                 &                            & 28.36 $\pm$ 0.09      & 2.87 $\pm$         0.25 \\
\hline
 Sun spot number                    & {\it 42178.78}      &     31.35         &                &                  &                 &                           & 31.35 $\pm$ 0.19      &  -   \\     
\hline
 \end{tabular}
 \end{table*}

\subsection{Comparison with previous work}


    It is not straightforward to determine the rotation period from a periodogram analysis. For the Sun, we usually
   take advantage of the fact that we already know the value of the rotation period we are searching for \citep[see][]{1998ApJ...493..494H}. 
   An early attempt to measure the solar surface differential rotation of the Sun as a star was made by
   \cite{1995SoPh..159...53D} based on the periodogram analysis of Ca~{\sc ii} K disc-integrated data. The authors reproduce the expected
   behaviour of the rotation period with time (with periods ranging from 22.8 to 28.0 days) with an abrupt
   jump in period in the transition between cycles 21 and 22. They also studied the pooled variance, finding
   that the evolution of active regions contribute to the total variance of the disc-integrated Ca~{\sc ii} K line at time scales from 20 to 400 days,
   which is a time scale compatible, although slightly longer time scale, than our results, see Figure~\ref{pooled_variance}. 

 \cite{1997A&A...322..835H} performed a wavelet analysis of disc-integrated solar Ca~{\sc ii} K line core emission measurements.
    The authors found the solar synodic rotation with a maximum power at a period of 28.3 days
    corresponding to the solar rotation at latitude 30$^{\circ}$. They also found a second (less dominant)
    period at 26.8 days as well as a third peak at 30.2 days related to the existence of active regions
    at 5$^{\circ}$ and 50$^{\circ}$, respectively.
    Our data, see Figure~\ref{activity_indices}, shows a period of   27.31 days as the dominant one, while secondary peaks at 28.62 and $\sim$ 26.30 days
    also appear.
    Using the known relationship between solar rotation rate and latitude it is possible to estimate the latitude of active regions
    at which our peaks would correspond.  
    There is a large amount of determinations of the solar rotation rate which depend on the methods and the type of data used,
    i.e., the type of feature used as a tracer, for a review see \cite{2000SoPh..191...47B}.
    It is known that chromospheric indicators provide higher rotation rates than those based on photospheric disc observations.
    This can be easily shown by comparing the equatorial rotation rates provided in Table~2 (based on measurements of
    solar features) and Table~1 (spectroscopic measurements) provided by \cite{2000SoPh..191...47B}.
    It is also clear from Table~\ref{table_sli_analysis} that our assumed rotation periods are more similar to the
    values derived from Table~1 of \cite{2000SoPh..191...47B}.
    \cite{1990ApJ...351..309S} derived three relationships, one related to magnetic features, a second one
    that refers to the rotation of supergranules
     (which are found to rotate faster than magnetic structures), and a third spectroscopic 
    (i.e., based on doppler measurements of spectral features)	relation.
    Using the coefficients provided by the latest relationship, our peaks would correspond to active regions located between the
    equator and  30$^{\circ}$.

    We have checked whether this could be the case by analysing the available data on active regions and
    solar spots at the Heliophysics Feature Catalogue\footnote{http://voparis-helio.obspm.fr/hfc-gui/}.
    Given the dates of our observations, the beginning of our time series should correspond to
    a fast rotation rate, while the end to a low rotation rate. 
    At the beginning of our observations, active regions were located at latitudes between -30$^{\circ}$ and 20$^{\circ}$ and
    sunspots were distributed into two regions between -10$^{\circ}$ and -20$^{\circ}$ and between 10$^{\circ}$ and 20$^{\circ}$.
    At the end of our observing window, active regions are mostly located between -10$^{\circ}$ and +20$^{\circ}$,
    while spots are seen into three groups at -10$^{\circ}$, 5$^{\circ}$, and 15$^{\circ}$, respectively. 
    We conclude that the presence of active regions located between the
    equator and  30$^{\circ}$ are indeed a plausible explanation for the observed periods.
    We also see that there seems to be no significant variation in the maximum latitude of active regions
    during our observation window.

    In their wavelet analysis, \cite{1997A&A...322..835H} found three ``bands'' of periods at around 26, 28, and 30 days.
    This is consistent with our sliding periodogram shown in Figure~\ref{peridogramas_hbeta_900} where 
    we see three main periods at around 26, 28, and 30 days. 
    A fourth period beginning at $\sim$ 32 days is also visible. 

    More recently \cite{2012ApJ...761...11B} measured 
    disc-integrated Ca~{\sc ii} K line  and observed periods in the range 27.7 - 28 days. 
    They find most of the power spectral density concentrated into a narrow band, whose central value varies
    in the range 26.3 to 28.6 days, consistent with the migration of active regions from latitudes of $\sim$  30-35$^{\circ}$ to 
    the equator. This result is consistent with the findings from our work. 
    It is worth noticing that while these authors found a single narrow band in their spectral analysis, the results from our power spectra are ``split'' into several
    main bands.  \cite{2012ApJ...761...11B} noted that while parameters defined in terms of wavelength separation
    show a spectrum with a single predominant peak, intensity-related parameters (like our measured indexes) show a power
    spectrum with a more complex pattern. Further differences can arise from the spectral estimator 
    used by these authors (the maximum entropy method) 
    that seems to produce less complex spectral patterns than the generalised Lomb-Scargle periodogram used here.

    Other attempts to measure the solar differential rotation have been performed
    using a combination of UV and optical lines using UARS SOLSTICE data
    \citep{2002AN....323..123H,2002A&A...388..540H} but no clear signatures
    of differential rotation were detected.
     We note that this result refers to integrated light.
    Using Sun-as-a star radial velocities
    measured during an eclipse,  \cite{2015PASJ...67...10T}
    tried to determine the coefficients of the solar rotation law\footnote{$\omega$ = A + B$\sin$$^{\rm 2}$$\psi$,
    where $\psi$ is the latitude.} 
    finding a relationship between the coefficients A, and B. However, an individual estimate 
    of the parameters A and B were not possible.



   \cite{1969SoPh....9..448L} noted an increase of the order of 8\% of the angular velocity measured in the chromospheric H$\alpha$ line
   with respect to other photospheric lines. 
   Regarding the activity indexes measured in this work,
   Na~{\sc i} D$_{\rm 1}$ D$_{\rm 2}$ is known to form in the upper photosphere and
   lower chromosphere, H$\alpha$, H$\beta$, and Ca~{\sc ii} H \& K in the middle cromosphere, while He~{\sc i} D$_{\rm 3}$
   is formed in the upper chromosphere \citep[e.g.][]{2000A&AS..146..103M}.
   No clear relationship between formation height and rotation rate is seen in our data. This conclusion is based on three facts:
   i) the rotation period derived from the Na~{\sc i} D$_{\rm 1}$ D$_{\rm 2}$ is similar to the one from 
   H$\beta$;
   ii)  H$\alpha$ show a higher period (between 2.5 and  7\%) than Ca~{\sc ii} and H$\beta$; and
   iii) as discussed above the rotation period found for the  He~{\sc i} index is less significant. 
   Nevertheless, we acknowledge that further studies should be performed, in particular the analysis of photospheric
   and coronal lines as well as a better knowledge of the lines' formation height would help to clarify this issue.  
   We  also tried to plot the assumed rotation period as a function of the line's excitation potential but no clear
   trend was evident. We finally checked for a tendency with the quantum number within the Balmer series,
   where H$\alpha$ show a higher period than H$\beta$ and both lines provide higher rotation levels than
   H$\epsilon$. Unfortunately, we were not able to detect the rotation period in H$\delta$ and the rotation period
   of H$\gamma$ does not follow this tendency.



\subsection{Correlation between activity indexes}
\label{activity_activity_correlations}

\begin{figure*}[!htb]
\centering
\includegraphics[angle=90,scale=0.80]{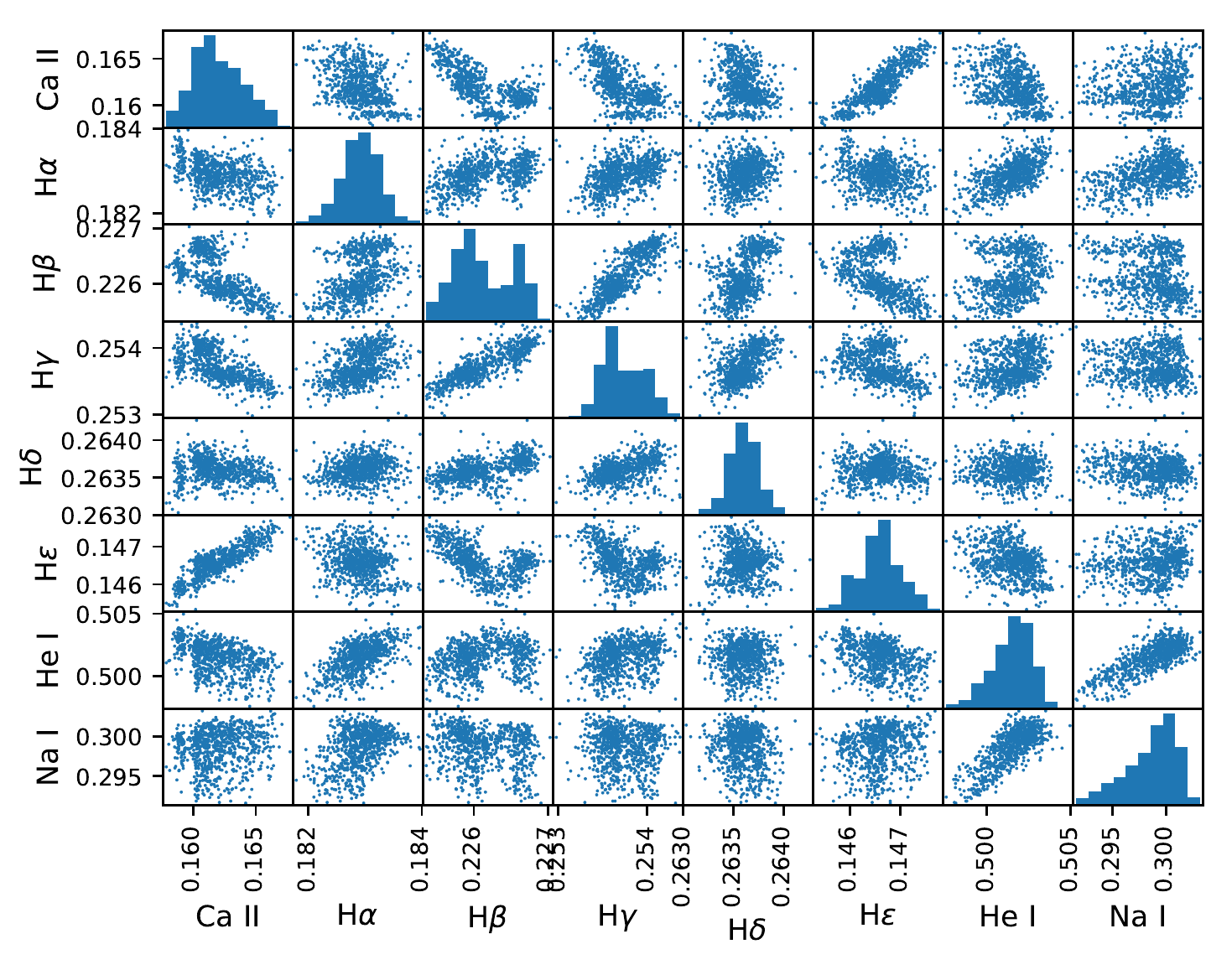}
\caption{ Correlation plot between the different activity indexes.
The diagonal panels show the histogram of each index. 
}
\label{activity_indices_vs_calcio}
\end{figure*}

 We also investigated the correlations between the different activity indexes
 described in Table~\ref{table_act_index}. Using the code MCSpearman\footnote{https://github.com/PACurran/MCSpearman}
 \citep{2014arXiv1411.3816C} we calculated correlations using a bootstrap Monte Carlo by adding 
  a Gaussian random variate with standard deviation equal to the measurement uncertainty to each data point. 
 For each pair of activity indexes, the corresponding Spearman coefficient correlation, $\rho$, and 
  its z-score\footnote{
   The  z-score measures the degree of significance of the correlation. 
A z-score of z approximately corresponds to a Gaussian significance of the correlation of $\sigma \sim z$.}
 were determined from  10$^{\rm 4}$ synthetic datasets. Figure~\ref{activity_indices_vs_calcio} shows the
 correlations between the different activity indexes, and the results from the statistical tests are provided in
 Table~\ref{table_act_act_corr}.  We note that the analysis presented in this section was made on the
 original time series (i.e., without any prewhitening applied).

 It can be seen that H$\alpha$, H$\beta$, H$\gamma$, and He~{\sc i} show a significant
 anticorrelation with the $S$ index 
 although the
 correlation coefficients are rather modest: 0.3-0.5.
 The H$\epsilon$ line, however, shows a positive correlation. 
 H$\delta$, and Na~{\sc i} 
 show no statistically significant  correlation with the $S$ index. 

 Table~\ref{table_act_act_corr} also reveals that the strongest correlation is typically with
 the closest index in wavelength (i.e, Ca~{\sc ii} with H$\epsilon$, H$\alpha$ with He~{\sc i},
 Na~{\sc i} with ~He{\sc i}, and H$\beta$ with H$\gamma$). While this could be an 
 instrumental effect (scattered light as a function of wavelength), if real,
 it could indicate a dependence on the formation height
 of the different indexes.

\begin{table}
\centering
\caption{Correlations between the activity indexes as defined in the text.} 
\label{table_act_act_corr}
\begin{tabular}{llrr}
\hline
X-index & Y-Index        &    $\rho$             &     z-score     \\
\hline
S            &  H$\alpha$        &    -0.3865 $\pm$       0.0320 &       4.925 $\pm$        0.455 \\
S            &  H$\beta$         &    -0.6686 $\pm$       0.0195 &       9.760 $\pm$        0.425 \\
S            &  H$\gamma$        &  -0.4619 $\pm$       0.0280 &       6.036 $\pm$        0.429 \\
S            &  H$\delta$        &    -0.0718 $\pm$       0.0349 &       0.869 $\pm$        0.423 \\
S            &  H$\epsilon$      &  0.6205 $\pm$       0.0238 &        8.768 $\pm$        0.467 \\
S            &  He~{\sc i}       &  -0.3921 $\pm$       0.0298 &       5.004 $\pm$        0.425 \\
S            &  Na~{\sc i}       &   0.1582 $\pm$       0.0322 &        1.927 $\pm$        0.398 \\
\hline
H$\alpha$    &  H$\beta$         &       0.4110 $\pm$       0.0302 &        5.277 $\pm$        0.439 \\
H$\alpha$    &  H$\gamma$        &       0.3088 $\pm$       0.0325 &        3.856 $\pm$        0.435 \\
H$\alpha$    &  H$\delta$        &       0.0806 $\pm$       0.0349 &        0.975 $\pm$        0.424 \\
H$\alpha$    &  H$\epsilon$      &      -0.1702 $\pm$       0.0353 &        2.076 $\pm$        0.440 \\
H$\alpha$    &  He~{\sc i}       &       0.5519 $\pm$       0.0261 &        7.502 $\pm$        0.455 \\
H$\alpha$    &  Na~{\sc i}       &       0.3377 $\pm$       0.0311 &        4.246 $\pm$        0.424 \\
\hline
H$\beta$     &  H$\gamma$        &       0.5911 $\pm$       0.0231 &        8.205 $\pm$        0.429 \\
H$\beta$     &  H$\delta$        &       0.2051 $\pm$       0.0331 &        2.514 $\pm$        0.417 \\
H$\beta$     &  H$\epsilon$      &      -0.3639 $\pm$       0.0306 &        4.606 $\pm$        0.426 \\
H$\beta$     &  He~{\sc i}       &       0.2560 $\pm$       0.0329 &        3.162 $\pm$        0.426 \\
H$\beta$     &  Na~{\sc i}       &      -0.1618 $\pm$       0.0349 &        1.971 $\pm$        0.433 \\
\hline
H$\gamma$    &  H$\delta$        &       0.1504 $\pm$       0.0342 &        1.831 $\pm$        0.423 \\
H$\gamma$    &  H$\epsilon$      &      -0.2168 $\pm$       0.0334 &        2.661 $\pm$        0.423 \\
H$\gamma$    &  He~{\sc i}       &       0.2270 $\pm$       0.0336 &        2.791 $\pm$        0.428 \\
H$\gamma$    &  Na~{\sc i}       &      -0.0524 $\pm$       0.0350 &        0.634 $\pm$        0.424 \\
\hline
H$\delta$    &  H$\epsilon$      &      -0.0120 $\pm$       0.0356 &        0.144 $\pm$        0.430 \\
H$\delta$    &  He~{\sc i}       &       0.0092 $\pm$       0.0352 &        0.111 $\pm$        0.426 \\
H$\delta$    &  Na~{\sc i}       &      -0.0400 $\pm$       0.0349 &        0.484 $\pm$        0.423 \\
\hline
H$\epsilon$  &  He~{\sc i}       &      -0.2611 $\pm$       0.0333 &        3.229 $\pm$        0.432 \\
H$\epsilon$  &  Na~{\sc i}       &       0.1386 $\pm$       0.0339 &        1.685 $\pm$        0.418 \\
\hline
He~{\sc i}   &  Na~{\sc i}       &       0.6926 $\pm$       0.0200 &       10.302 $\pm$        0.463 \\     
\hline
\end{tabular}
\end{table}

 As mentioned, one result from Figure~\ref{activity_indices_vs_calcio} and the analysis presented in Table~\ref{table_act_act_corr} is that $S$ emission and H$\alpha$
 seem to anticorrelate. This might be an unexpected behaviour as both quantities are well known to correlate with the solar cycle
 \citep[e.g.][]{2007ApJ...657.1137L,2009A&A...501.1103M} and a good correlation between the $S$ index and H$\alpha$ 
 has been found for other stars \citep{2016A&A...589A.135F}.
 However, it should be noted that these first works are based on nearly thirty and twenty years of observations, while the short-term correlation
 between $S$  and H$\alpha$ and its relationship with the level of activity is still not well understood 
 \citep[e.g.][]{1989ApJ...345..536G,1990ApJS...72..191S,1990ApJS...74..891R,1995A&AS..109..135M}.
 It is also worth noticing that a wide variety of correlations between -1 and +1 have been found in other stars with spectral types between F6
 and M5 \citep{2007A&A...469..309C,2011A&A...534A..30G,2014A&A...566A..66G}.
 It has also been shown that stars exhibiting a positive correlation show a tendency to be more active
 and that negative correlations are more present among higher metallicity stars \citep{2014A&A...566A..66G}.
 In particular, 
 we note that these authors found that
 stars more active than  $\log$R'$_{\rm HK}$ = -4.7 showed positive correlations, while the median solar value is -4.907 \citep{2008ApJ...687.1264M}. 
 Although the Sun is not a metal-rich star, we note that these authors found negative correlations even in stars with [Fe/H] $\sim$ -0.15. 

\begin{figure}[!htb]
\centering
\includegraphics[scale=0.45]{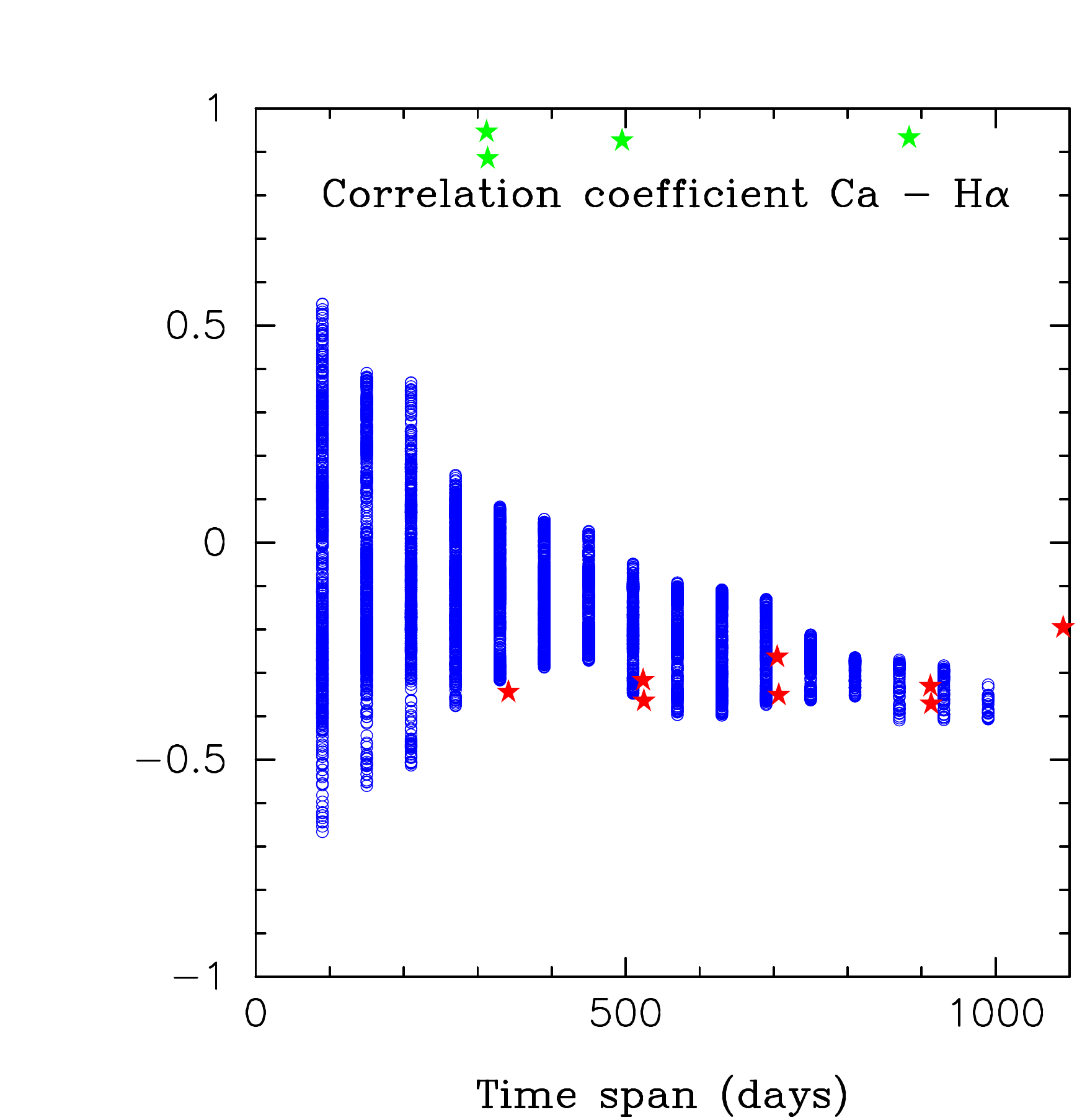}  
\caption{ Correlations between the $S$ and H$\alpha$ activity indexes as a function of the time span.
Red and green stars denotes the lower and upper boundary of the data shown in \citet[][Figure~3]{2009A&A...501.1103M}.
}
\label{comparison_with_meunier}
\end{figure}

 In order to test whether the negative correlation coefficient  between $S$ and H$\alpha$ is related to the time scale of the observations we
 study how the correlation depends on the timescale of the observations. 
 This is shown in Fig.~\ref{comparison_with_meunier}. The correlation between $S$ and H$\alpha$ is computed for different time intervals
 and different starting times in the series (i.e. different cycle phases). For each time interval, individual
 circles correspond to a different starting time in the series. 
 We overplot with red stars the lower boundary of the data presented in \citet[][Figure~3]{2009A&A...501.1103M} while
 with green stars we show the upper boundary.
 The figure shows that shorter time lags lead to more negative correlation coefficients.
 In particular, negative correlation coefficients as the ones obtained in this work are consistent with
 the findings of \cite{2009A&A...501.1103M}. However, unlike these authors we are not able to obtain
 correlations close to one in any of the inspected time lags.
 
 We conclude that correlation coefficients as low as $\sim$ -0.30 can be obtained for short timescales in agreement
 with previous results. However, the question of why we do not get any value close to one remains open.
 At this point two explanations are possible: either the two lines show a different response to increasing activity or
 the correlation between $S$ and H$\alpha$ depends on
 the level of activity. We should note that the data by \cite{2009A&A...501.1103M} covers almost 30 years of observations,
 so they can really test different parts of the solar cycle in each temporal window. However, our observations do cover
 $\sim$ 2.9 yrs so even if we choose different starting times
  we  do not
 trace different  parts of the  cycle, only  the end of  the decreasing
 phase and around the minimum.
 This is  specially evident when large windows are considered, for example, if a window including 900 days is explored, starting dates
 can not be separated by more than 32 days, so we are essentially in the same phase of the cycle.

 Regarding the behaviour of  Ca~{\sc ii} and H$\alpha$ with activity,
 it is well known that the H$\alpha$ line is radiation-dominated and it first shows a deeper absorption profile as activity increases until
 the electron density is high enough and the line becomes collisionally excited, then increasing the emission at its core. 
 On the other hand,  chromospheric Ca~{\sc ii} H \& K emission lines are collisionally dominated, thus the radiated flux steadily increases with electron pressure
 \citep{1979ApJ...234..579C,1987ApJ...323..316C,1989A&A...219..239R,1995A&A...294..773H,1997A&A...326.1143H,2017A&A...598A..28S}.
 On the other hand, the dependency of the correlation between $S$ and H$\alpha$ with the level of activity is discussed at length in Section~\ref{disc_section}.

  Some degree of correlation between Ca~{\sc ii} and H$\alpha$ is however visible in our data.
  This can be seen in Figure~\ref{activity_indices}. At the beginning of the time series, the S index shows a variation due to the crossing
  of active regions (we have checked using solar images that several
  active regions were effectively crossing the solar disc during these days).
  This is also visible in  H$\alpha$, i.e., there seems to be a correspondence between peaks in both indexes,
  although the higher noise level in the H$\alpha$ index makes the comparison difficult. 
  Another hint of correspondence between peaks in both indexes can be seen in the middle panel of Figure~\ref{activity_indices}
  which shows a zoom around BJD = 2458000 days. 
  We should note that we do not have a clear reason of why the  H$\alpha$ time series is noisier. We expeculate that it could be related to contamination by
  some non-corrected nearby telluric line. Whether the higher noise level in the H$\alpha$ data might contribute to hide a clearer S index - H$\alpha$ correlation
  is unclear. As mentioned, there are other factors that should be considered.

 The results from the  He~{\sc i} and Na~{\sc i} indexes also deserve further comments.
 In particular, we should note that clear cyclic variations in the Na~{\sc i} index have been found for the Sun
 \citep{1998ApJ...493..494H,2007ApJ...657.1137L} and a clear correlation between Na~{\sc i} and the $S$ index
 in low activity M dwarfs has been reported \citep{2011A&A...534A..30G}.
 However, from our results no statistical significant correlation is found between these two indexes.
 Regarding He~{\sc i} measurements, we find a statistical significant anticorrelation with the $S$ index, while
 He~{\sc i} observations of the 1083 nm line 
 have been shown to show variations correlated with the solar cycle \citep{2007ApJ...657.1137L}.
 We should note that the He~{\sc i} line at 1083 nm has been shown to be negatively correlated with activity,
 i.e., it becomes deeper when the Sun is more active \citep{1995ApJ...439..405A}, in line with the findings of our work. 
 Furthermore,  \cite{2011A&A...534A..30G} found little variation of  He~{\sc i} index
 (measured in the same line than our work)
 as well as a low correlation
 with activity, although their results refers to M dwarfs. 
 
 We believe that as might happens with the H$\alpha$ the lack of a clear (positive) correlation of these indexes with
 the $S$ index is due to the relatively small temporal window covered in this work as well as a possible dependence
 of the correlations on the level of activity,  note that our observations cover the decreasing phase and
 the minimum of the solar cycle. Furthermore, these indexes also show higher levels of noise.

 We finally caution that the presence of etalon ripples on the HARPS-N solar data has been reported (see Thompson et al., in prep.).
 In order to understand whether they can effect  the correlations between activity indexes  
 we took several random spectra and computed the GLS periodogram of 50 and 100 \AA \space width regions centred around the activity indexes,
 finding no evidence of ripples in these regions. At the time of writing, we are not aware of any other possible anomaly in the data that
 might affect our measurements. 

  A future analysis of the solar images covering the same time span analysed in this work would be helpful to clarify the trends
  between  the different activity indicators found in this work as well as the different significance levels of 
  the periods found in these indices.

\subsection{Correlation activity indexes - CCF parameters}
\label{activity_ccf_correlations}

\begin{figure}[!htb]
\centering
\includegraphics[scale=0.45]{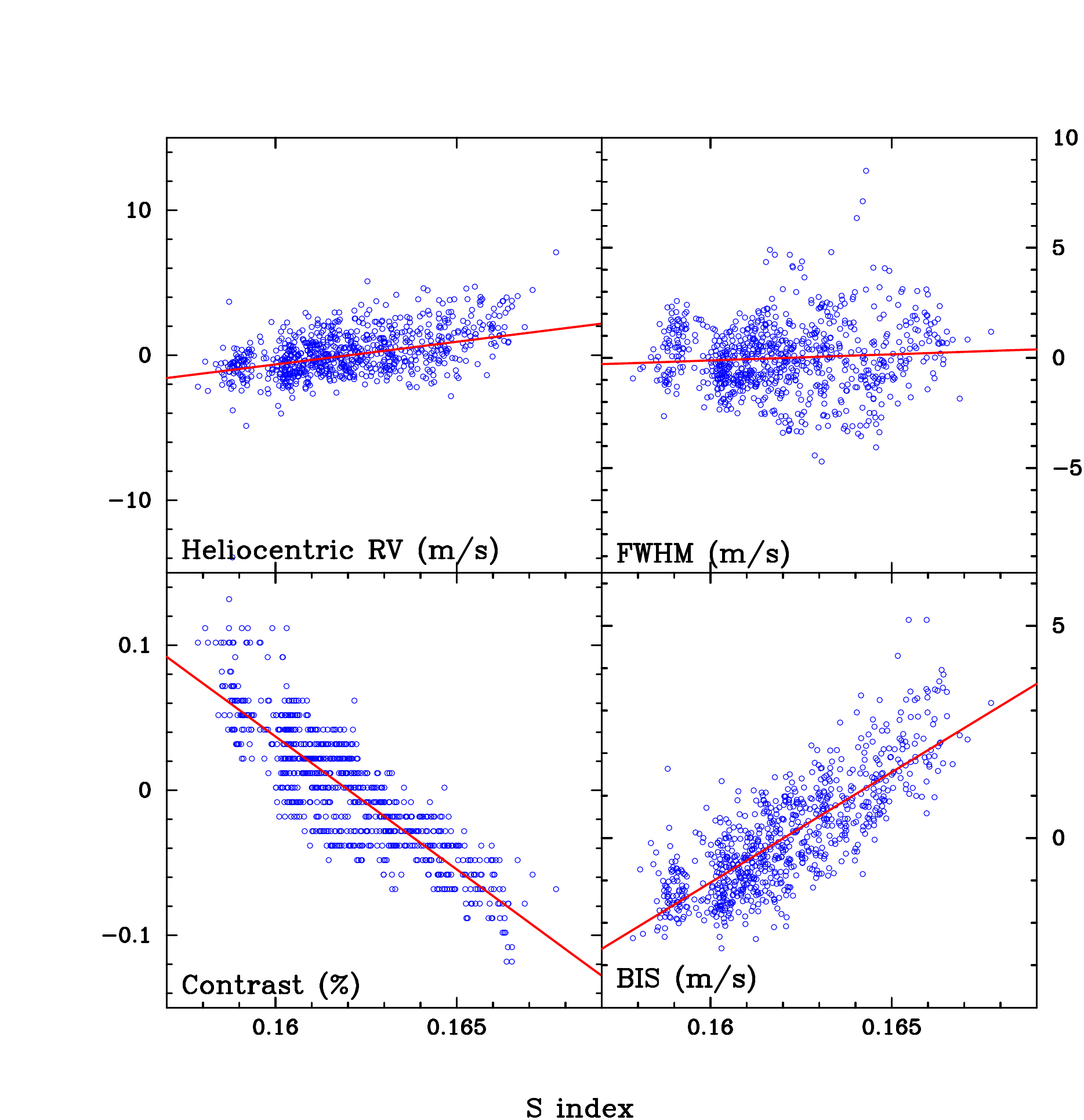}
\caption{ 
CCF parameters vs. the Ca~{\sc ii} H \& K index. Best linear fits are shown in red lines.
}
\label{ccf_indices_vs_calcio}
\end{figure}

 Possible correlations between the RVs and several CCF asymmetry diagnostics were also investigated. 
 We considered the CCF width (FWHM), the bisector  span \citep[][BIS]{2001A&A...379..279Q}, and its contrast. 
 The  heliocentric radial velocity (as described in
 Sect.~\ref{observations_sect}) 
 was also considered.
 In addition,  FWHM values were also corrected
 to the sidereal frame, see \cite{2019MNRAS.tmp.1180C}. 
 Spearman correlation coefficients, $\rho$ and z-scores for these correlations are given in Table~\ref{table_act_ccf_corr},
 and Figure~\ref{ccf_indices_vs_calcio} shows the correlations for the Ca~{\sc ii} S index. 
  The analysis presented in this section corresponds to the
  original time series (i.e., without any prewhitening applied).

 We search for significant correlations using a z-score larger than 3.0 in absolute value, the usual threshold for statistical significance.
 These are shown in bold face in Table~\ref{table_act_ccf_corr}.
 The Ca~{\sc ii} index shows a significant correlation with the bisector
 and the heliocentric RV  values while it shows an anticorrelation with the
 contrast. 
 A similar behaviour is seen for the H$\epsilon$ line.
 Also  H$\beta$ shows the same correlations but in all cases
 with the opposite sign. Additional significant correlations are found
 between the H$\alpha$ index and the CCF contrast and between the
 H$\gamma$ and Na~{\sc i}  indexes and the contrast and bisector (with opposite signs).
 Furthermore, the He~{\sc i} index also  correlates with the CCF contrast. 
 No correlations are found between the  H$\delta$ 
 index and the CCF parameters.

 It is important to note the positive correlation between the heliocentric
  radial velocity and the activity index $S$.
 Our derived Spearman's correlation coefficient of 0.40 $\pm$ 0.03 (with a 5.1$\sigma$
 confidence level) is in agreement with the value of 0.357 found by \cite{2016A&A...587A.103L}.
 The rms of the RV variations is 4.9 ms$^{\rm -1}$, also in agreement with these authors. 
 Time delays between the FWHM and BIS values with the RV variations have been predicted by models \citep{2014ApJ...796..132D}
 and found in the HARPS-N solar data by \cite{2019MNRAS.tmp.1180C}. 
 We searched for similar time delays between the $S$ index and RV variations
 but no evidence was found.
%
 The correlation between the emission in the Ca~{\sc ii} line and the radial velocity supports
 the idea of the quenching of local convective motions in active regions being the main effect on the Sun as a star radial velocities, 
 as predicted by \cite{2010A&A...519A..66M} and \cite{2013A&A...551A.101M}.
 This has been observationally confirmed by \cite{2016MNRAS.457.3637H} and \cite{2019ApJ...874..107M} 
 On the other hand, the effect of inhomogeneities in the solar surface (such as dark
 spots or bright faculae) can be either positive or negative depending on the
 filling factors and the location of active regions (receding or approaching half of the solar
 disc), leading to a weaker 
 $S$ index - RV correlation \citep{2010A&A...512A..38L,2010A&A...519A..66M,2010A&A...512A..39M,2014ApJ...796..132D}.
 
 Along these lines, \cite{2016MNRAS.457.3637H} searched for correlations between the RV variations of the Sun-as-a-star
 and the RV component due to the suppression of the convective blueshift with the
 $\log$(R'$_{\rm HK}$) values. 
 When considering the overall RV, they find a correlation coefficient of 0.18, i.e, significantly 
 lower than our value, while the  coefficient increases to 0.35 (more similar to our value) when only the RV due to the suppression of convective blueshift is taken into account.
 It is important to note that this comparison may be biased by the fact that the radial velocities 
 from this previous work were taken at a different time in the activity cycle
 (in particular the Sun was more active than in our observing season),
 so the fact that we obtain similar correlation coefficients to the \cite{2016MNRAS.457.3637H}
 values might be coincidental.
 We note that the data by \cite{2016MNRAS.457.3637H} were collected from September 29 to  December 7, 2011 and were spread over 38 nights for a total of 98 datapoints. Note also that the first part of the RV data suffered from an additional scatter owing  to the finite diameter of the asteroid 4/Vesta being comparable with the fiber diameter on the focal plane. 

\begin{table*}
\centering
\caption{Correlations between the different activity indexes and 
the CCF indexes. The correlations with a z-score larger than 3.0 are
shown in bold face.}
\label{table_act_ccf_corr}
\begin{tabular}{lcccc}
\hline
	   &  Heliocentric RV 		&	FWHM		&	CONTRAST  & BIS	 \\ 
 \hline
 \multicolumn{5}{c}{\bf Ca~{\sc ii} H \& K}\\
 \hline
$\rho$  &       0.452   $\pm$        0.029  &       0.048 $\pm$        0.035 &      -0.835 $\pm$        0.012 &       0.642 $\pm$        0.022 \\
z-score &   {\bf  5.890 $\pm$        0.440} &       0.585 $\pm$        0.424 &   {\bf   14.528 $\pm$        0.465} &  {\bf     9.198 $\pm$        0.452 }\\
 \hline
 \multicolumn{5}{c}{\bf H$\alpha$}\\
 \hline
$\rho$  &      -0.035 $\pm$        0.035 &             0.132 $\pm$        0.034 &       0.286 $\pm$        0.033 &      -0.173 $\pm$        0.034  \\
z-score &       0.425 $\pm$        0.419 &             1.610 $\pm$        0.417 &  {\bf  3.557 $\pm$       0.431} &       2.110 $\pm$        0.423  \\
 \hline
 \multicolumn{5}{c}{\bf H$\beta$}\\
 \hline
$\rho$  &      -0.248 $\pm$        0.034 &             0.129 $\pm$        0.035 &       0.592 $\pm$        0.023 &      -0.481 $\pm$        0.027  \\
z-score &   {\bf    3.059 $\pm$        0.437} &        1.562 $\pm$        0.425 &     {\bf  8.227 $\pm$        0.421} &  {\bf     6.330 $\pm$        0.431}  \\
 \hline
 \multicolumn{5}{c}{\bf H$\gamma$}\\ 
 \hline
$\rho$  &      -0.130 $\pm$        0.035 &            0.101 $\pm$        0.035 &       0.380 $\pm$        0.030 &      -0.280 $\pm$        0.032  \\
z-score &       1.577 $\pm$        0.428 &            1.224 $\pm$        0.426 &    {\bf   4.838 $\pm$        0.421} &   {\bf    3.478 $\pm$        0.421}  \\
 \hline
 \multicolumn{5}{c}{\bf H$\delta$}\\
 \hline
$\rho$  &       0.036 $\pm$        0.035 &             0.086 $\pm$        0.035 &       0.035 $\pm$        0.036 &      -0.063 $\pm$        0.035  \\
z-score &       0.436 $\pm$        0.428 &             1.047 $\pm$        0.423 &       0.424 $\pm$        0.429 &       0.763 $\pm$        0.427  \\
 \hline
\multicolumn{5}{c}{\bf H$\epsilon$}\\
 \hline
$\rho$  &       0.328 $\pm$        0.032 &            -0.018 $\pm$        0.035 &      -0.651 $\pm$        0.022 &       0.497 $\pm$        0.028  \\
z-score &    {\bf   4.111 $\pm$        0.436} &        0.217 $\pm$        0.423 &   {\bf    9.385 $\pm$        0.470} &     {\bf  6.585 $\pm$        0.442}  \\
 \hline
 \multicolumn{5}{c}{\bf He~{\sc i} D$_{\rm 3}$ }\\
 \hline
$\rho$  &      -0.106 $\pm$        0.035 &            0.066 $\pm$        0.035 &       0.320 $\pm$        0.032 &      -0.142 $\pm$        0.034  \\
z-score &       1.286 $\pm$        0.425 &            0.802 $\pm$        0.421 &     {\bf  4.011 $\pm$        0.426} &       1.721 $\pm$        0.413  \\
 \hline
\multicolumn{5}{c}{\bf Na~{\sc i} D$_{\rm 1}$ D$_{\rm 2}$ }\\
 \hline
$\rho$  &       0.185 $\pm$        0.035 &            0.019  $\pm$        0.035 &      -0.213 $\pm$        0.033 &       0.291 $\pm$        0.032  \\
z-score &       2.262 $\pm$        0.437 &            0.229  $\pm$        0.421 &       2.613 $\pm$        0.416 &     {\bf  3.617 $\pm$        0.421}  \\
\hline
\end{tabular}
\end{table*}

\begin{figure*}[!htb]
\centering
\begin{minipage}{0.33\linewidth}
\includegraphics[scale=0.425]{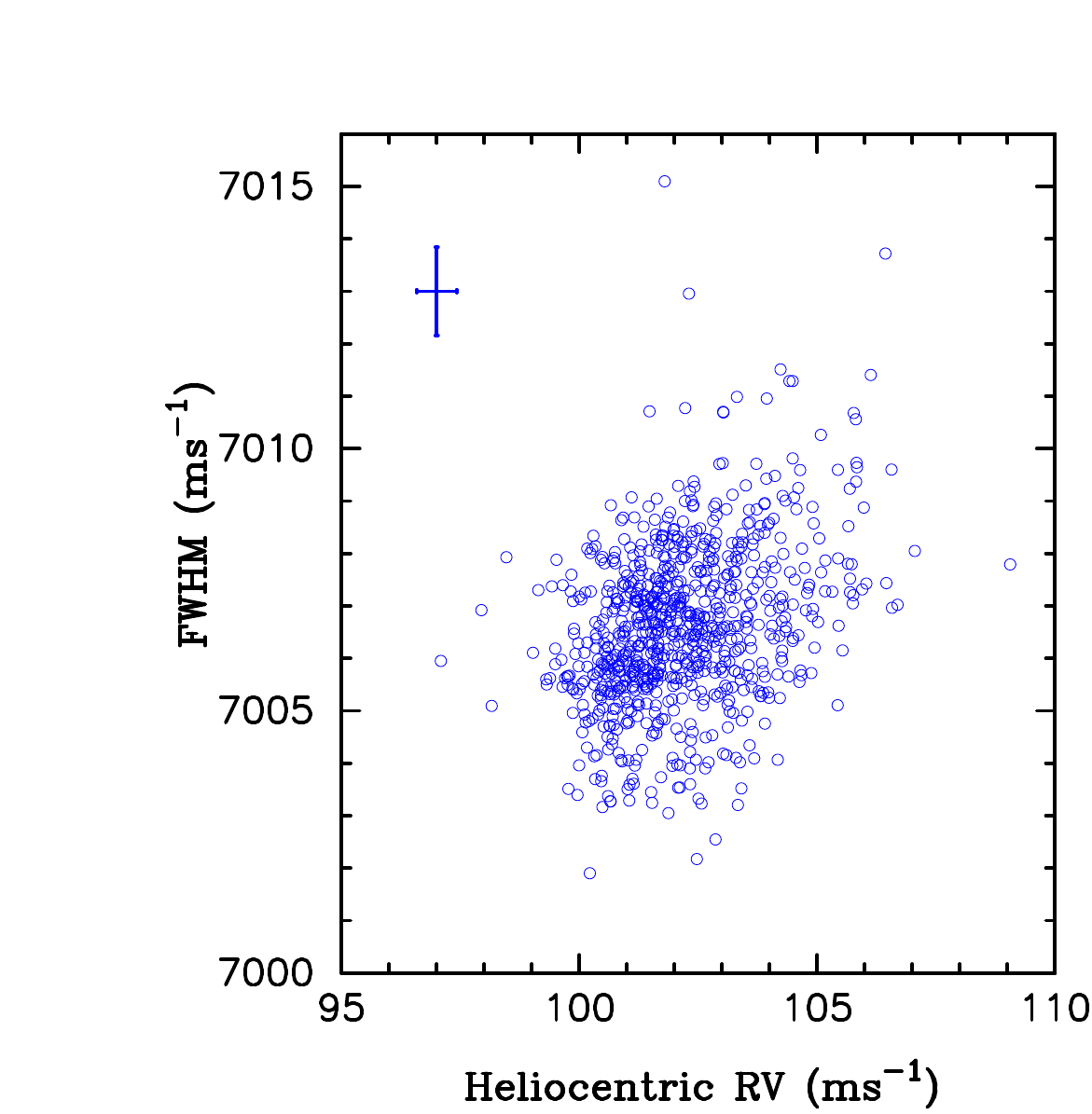}
\end{minipage}
\begin{minipage}{0.33\linewidth}
\includegraphics[scale=0.425]{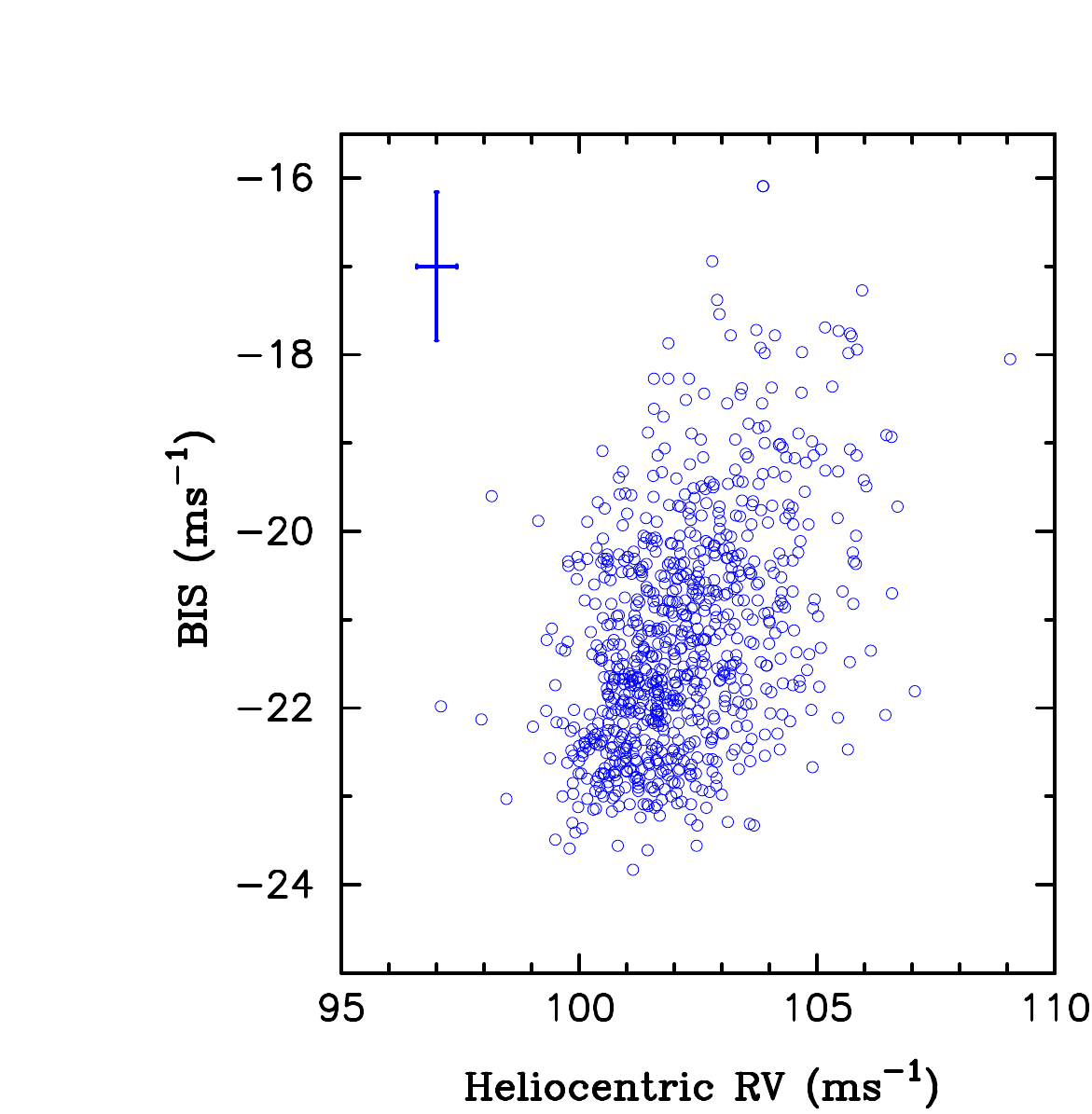}
\end{minipage}
\begin{minipage}{0.33\linewidth}
\includegraphics[scale=0.425]{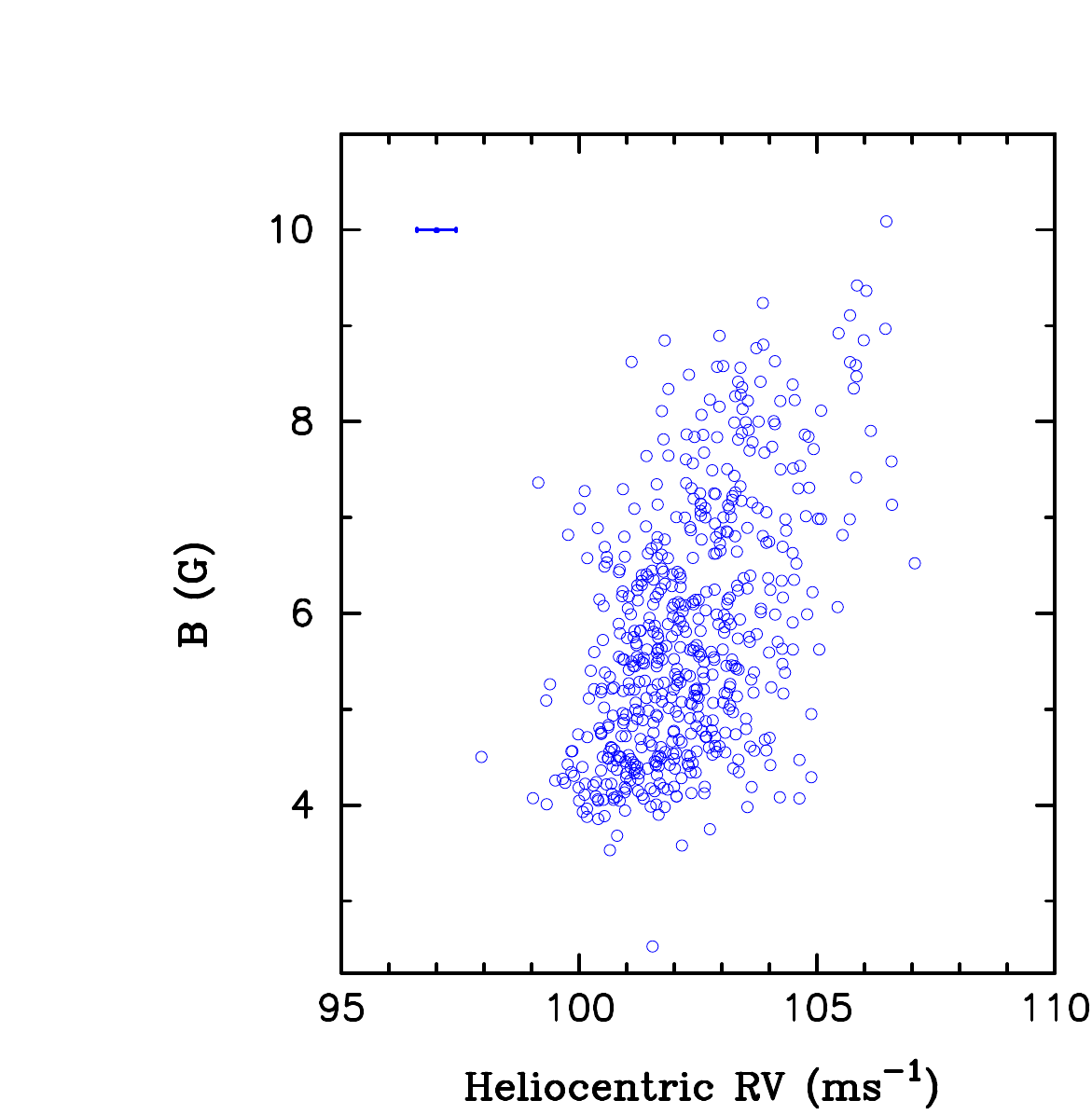}
\end{minipage}
\caption{
Sidereal FWHM (left),
bisector span (centre), and  
disc-integrated magnetic flux (right)
as a function of the heliocentric radial velocity.
}
\label{fwhm_bis_vrad}
\end{figure*}

 In a recent work \cite{2018A&A...616A.155L}, studied the correlations of different CCF indicators, in particular the FWHM and the BIS,
 with the radial velocity variations for a sample of 15 solar-type stars. 
 The authors found significant correlation in 13 (FWHM) and 27 (BIS) percent of the stars.
A search for similar correlations were performed in our data. The results are shown in Figure~\ref{fwhm_bis_vrad} where the FWHM 
 and BIS values are plotted against the heliocentric radial velocity.
 Our analysis reveals a significant positive correlation between the FWHM and the BIS with the heliocentric radial velocity.
 For the FWHM the Spearman's rank coefficient is 0.267 $\pm$ 0.033 with a z-score value of 3.310 $\pm$ 0.429. 
 When considering the BIS values, we obtain a Spearman's rank coefficient of 0.325 $\pm$ 0.032 and a z-score value of
 4.075 $\pm$ 0.435.

 Finally, we studied the relationship between the mean magnetic flux 
 derived from SOLIS/VSM full-disc line-of-sight magnetograms which correspond to the core of the photospheric 630.15 nm Fe line 
 \citep{2002ESASP.505...15J}\footnote{https://solis.nso.edu/0/vsm/vsm\_mnfield.html}
and the heliocentric  radial velocity. The corresponding plot is shown in the right panel of Figure~\ref{fwhm_bis_vrad}.
 A significant correlation is found with a Spearman's rank coefficient of 0.463 $\pm$ 0.04 and a significance z-score of
 5.026 $\pm$ 0.452. 
 We note that our derived Spearman's coefficient is significantly larger than the 0.131 found by 
 \cite{2016A&A...587A.103L}, and is in agreement with the 0.58 value found by 
 \cite{2016MNRAS.457.3637H}. It is lower than the 0.87 value found when only the radial velocity
 contributions from convective blueshift (the factor that dominates the radial velocity of the Sun as a star) is considered
 \citep{2016MNRAS.457.3637H}.

\section{Discussion}
\label{disc_section}

 The temporal variation of the correlations coefficients described in Sect.~\ref{activity_activity_correlations} and ~\ref{activity_ccf_correlations} was also studied.
 As in Sect.~\ref{stacked_periodograms}, correlations tests were performed considering a large window of  600 days period
 of observations every two days. 
 We focus on two correlations: 
 those between  H${\alpha}$
 and the  Ca~{\sc ii} H \& K indexes and 
 between the heliocentric  RV and the Ca~{\sc ii} H \& K index. 
 The top of Fig.~\ref{correlations_all_plots} shows variations in the correlation of these
 quantities changes with time.

\subsection{Understanding the Ca - H$\alpha$ correlation}

 In order to understand the correlations between the emission in the Ca~{\sc ii} H \& K and
 the H$\alpha$ lines, in each temporal window the slope of the S versus H$\alpha$ index relationship
 was computed. The slopes are then plotted versus the mean time of the temporal window as well as versus the mean
 values of  different observables related
 to the evolution of magnetic structures in the solar atmosphere, Figure~\ref{correlations_all_plots} (left). 
 These observables are 
 i) the daily total sunspot number by the Royal Observatory of Belgium\footnote{http://www.sidc.be/silso/datafiles};
 ii) the total solar irradiance data from the University of Colorado\footnote{http://lasp.colorado.edu/home/sorce/data/tsi-data/};
 and iii) the solar f10.7 radio flux from the University of
 Colorado\footnote{http://lasp.colorado.edu/lisird/data/penticton\_radio\_flux\_observed/}.
 While other possible diagnostics are available, they do not cover the dates of our 
 HARPS-N observations. 

  In this analysis we use the original time series of the indexes. 
 The figure reveals a tendency of  less negative slopes as we move towards larger sun spot numbers,
 irradiance, or radio flux values.
 We finally tested whether the derived slope correlates with these quantities.
 For this purpose we use the procedure described
 in Sect~\ref{activity_activity_correlations}.
 The results are given in Table~\ref{table_ca_ha_cor}.
 It shows the correlation coefficient and z-score between the slopes (of the
 $S$ index vs. H$\alpha$ index relationship) and the sun spot, irradiance, and radio flux values. 
 The results from this statistical test do not exclude that correlations are present but suggest
 that they are relatively weak.
 A dependence of the $S$ index - H$\alpha$ correlation with the level of activity has also been found in 
 FGK stars \citep{2014A&A...566A..66G} as well as in a sample of early-M dwarfs by \cite{2011A&A...534A..30G}.
 It is interesting to note that 
 \cite{2014A&A...566A..66G} also found this correlation to be dependent on the stellar metallicity suggesting that negative 
 correlations were present among higher metallicity stars, but as we are studying one single
 star we are not able to test this result.

\begin{table}
\centering
\caption{ Correlations between the slope of the $S$ index and H$\alpha$ relationship and  different 
observables of the evolution of magnetic structures in the solar atmosphere.} 
\label{table_ca_ha_cor}
\begin{tabular}{lrr}
\hline
            &    $\rho$             &     z-score     \\
\hline
 Sun spot   &     -0.2797 $\pm$       0.0582 &       1.788 $\pm$        0.393 \\
 Irradiance &     -0.2932 $\pm$       0.0582 &       1.880 $\pm$        0.396 \\
 Radio flux &     -0.4068 $\pm$       0.0501 &       2.623 $\pm$        0.364 \\
\hline
\end{tabular}
\end{table}

 It is well known that at short time scales the emission in the Ca~{\sc ii} H \& K and
 H$\alpha$ do not always correlate. 
 \cite{2009A&A...501.1103M} suggest that the presence of plages might lead to a correlation coefficient lower than 1
 as the surface covered by the H$\alpha$ plages is slightly smaller than the surface
 covered in Ca (the calcium lines are formed higher in the chromosphere), and that
 also the chromospheric network is more prominent in calcium. 
 The different contrast of dark filaments in Ca and H$\alpha$ might also decrease
 the correlation coefficient between Ca and H$\alpha$ or even lead to an
 anti-correlation, if strong filaments correlated with plages are present
 \citep{2009A&A...501.1103M}. Indeed, it could be that filaments are actually more prominent in all
 Balmer's lines, which might partly explain their anticorrelation with the 
 Ca index (see Table~\ref{table_act_act_corr}).

 Our results seem to be consistent with the scenario described above. At large sunspot
 numbers the emission in Ca and H$\alpha$ show
 less negative slopes suggesting
 different relative contributions of the bright plages and dark filaments when compared to time intervals with a lower level of activity.

\subsection{Understanding the Ca - RV correlation}

\begin{figure*}[!htb]
\centering
\begin{minipage}{0.49\linewidth}
\includegraphics[scale=0.50]{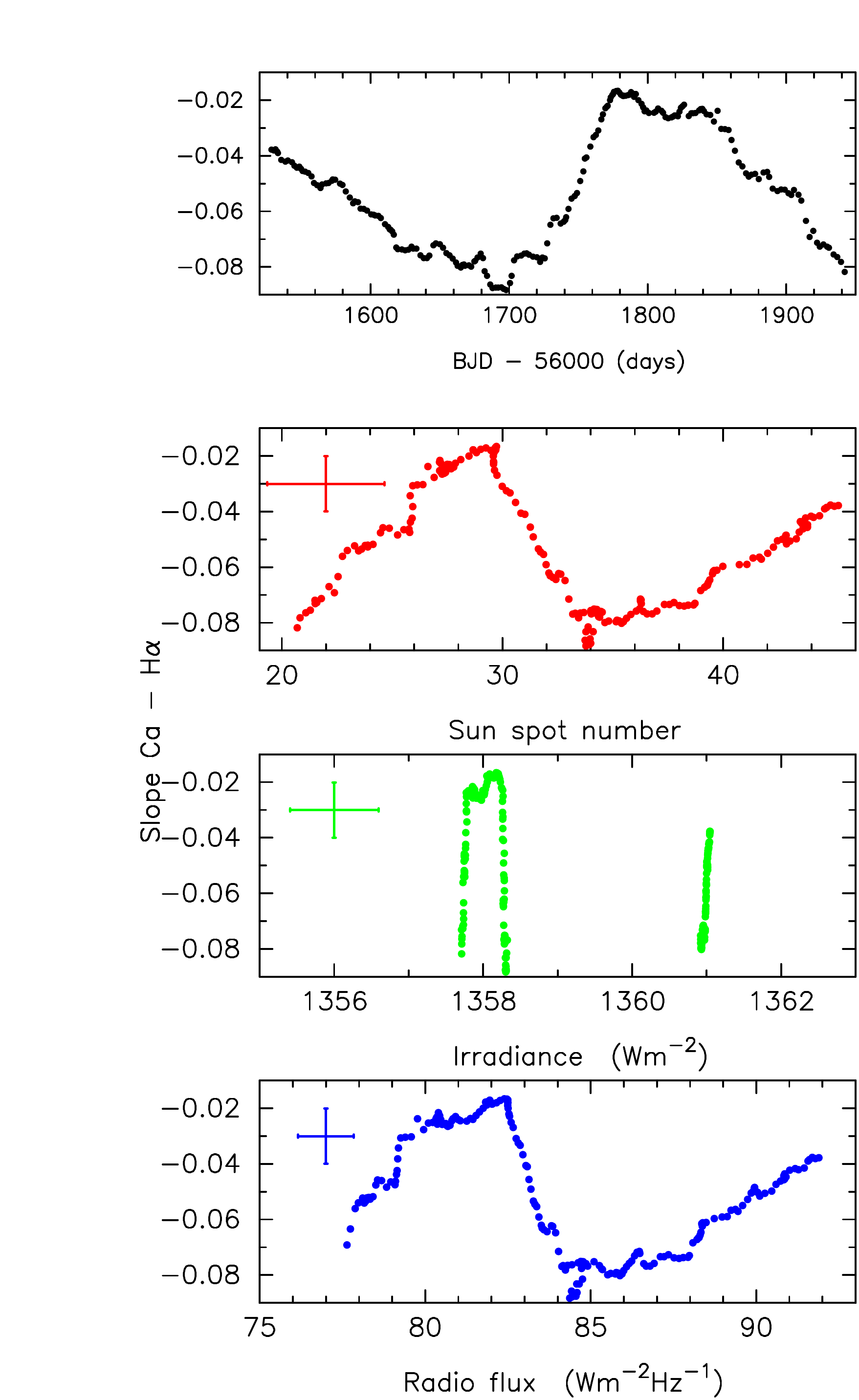}
\end{minipage}
\begin{minipage}{0.49\linewidth}
\includegraphics[scale=0.50]{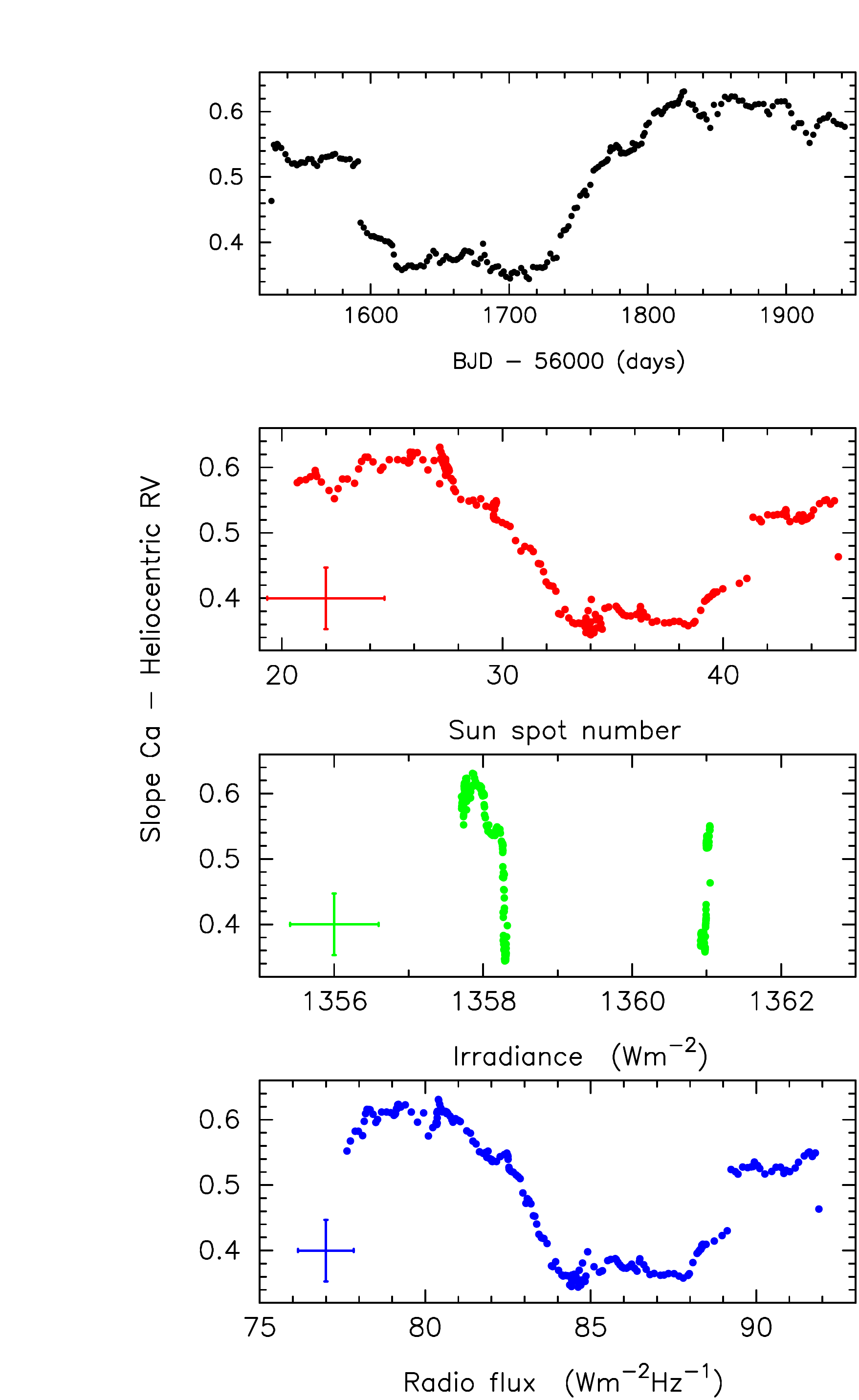}
\end{minipage}
\caption{ 
Slope of the relationship between Ca, and H$\alpha$ indexes (left), and
slope of the relationship between Ca, and the heliocentric RV (right)
as a function
of time, the sunspot index, the solar irradiance, and the solar radio flux.
}
\label{correlations_all_plots}
\end{figure*}

 We now search for correlations between the 
  emission in the Ca~{\sc ii} H \& K and the heliocentric  radial velocity.
 Figure~\ref{correlations_all_plots} (right) shows the slope 
 between the daily median S and the daily median heliocentric RV as a function of time
 and 
 sunspot indexes.

 A general tendency of a larger degree of correlation towards larger activity levels
 is present in the data, see Table~\ref{table_ca_vr_cor}, suggesting that a higher presence of surface inhomogeneities
 produces larger radial velocity variations.
 As mentioned, the correlation between the Ca~{\sc ii} H \& K and the heliocentric  radial velocity
 is due to the suppression of convective blueshift by magnetic areas, leading to
 a net redshift. Therefore on cycle timescales, as total magnetic areas increase and decrease,
 the associated redshift will do likewise \citep[e.g.][]{2000ApJ...534L.105S}. 

 It is worth noticing that the slope is not constant with time. That might be expected if larger filling fractions
 of magnetic regions effectively translate to larger radial velocity variations.
 Therefore, we conclude that there should be  non-magnetic effects that swap the magnetic variations
 at epochs of low activity, thus reducing the apparent correlation.

\begin{table}
\centering
\caption{ Correlations between the slope of the $S$ index and heliocentric radial velocity relationship and  different 
observables of the evolution of magnetic structures in the solar atmosphere.} 
\label{table_ca_vr_cor}
\begin{tabular}{lrr}
\hline
            &    $\rho$             &     z-score     \\
	    \hline

 Sun spot   &     -0.4992 $\pm$       0.0515 &       3.414 $\pm$        0.426 \\
 Irradiance &     -0.4051 $\pm$       0.0547 &       2.675 $\pm$        0.407 \\
 Radio flux &     -0.5050 $\pm$       0.0586 &       3.385 $\pm$        0.478 \\
	    \hline
	    \end{tabular}
	    \end{table}

\section{Conclusions}
\label{summary} 

 In this work a detailed analysis of the main optical activity indicators including the Ca II H \& K doublet, the
 Balmer lines, the  Na~{\sc i} D$_{\rm 1}$ D$_{\rm 2}$ doublet,  and the He~{\sc i} D$_{\rm 3}$ measured in the
 Sun-as-a-star observations has been presented. Nearly three years of data were analysed.

 The periodogram analysis of the data reveals the solar rotation period in almost all activity indicators with values ranging
 from 26.29 days to 31.23 days. The H$\delta$ line is the only exception in which the rotation period was not found,
 suggesting that either this line is less sensitive to activity or that its definition should be revised.
 Furthermore, there is no clear reason why it does not correlate with other Balmer lines.
 The pooled variance analysis reveals two main slope changes. The first one is at around 30 days corresponding to the solar rotation period.
 The second slope change occurs at $\sim$ 300 day,
 suggesting that the typical lifetime of complexes of activity dominated by plage and faculae is around 10 rotations periods, in agreement
 with previous results \citep[e.g.][]{1986SoPh..105..237C}.
 The study using sliding periodograms shows that
 the spectral power is usually split into several bands with periods ranging from 26 to 30 days. This effect 
 might be due to differential rotation 
 (migration of active regions between the solar equator and a latitude of $\sim$  30$^{\circ}$), active region evolution, or a mixture
 of both phenomena. 
 Again, no significant periods are found for the H$\delta$ line.

 We studied the correlations between the different activity indicators. Excluding H$\delta$ and the Na~{\sc ii} doublet, all the 
 indicators show a modest but statistically significant correlation with the $S$ index. In all cases, with the exception of H$\epsilon$, the sign of the
 correlation is negative.

 Significant positive correlations are found between the bisector and the heliocentric  radial velocity with the $S$ index and the
 H$\epsilon$ index, as well as an 
 anticorrelation with the contrast. The  H$\beta$ line show similar correlations, but
 always with an opposite sign. Additional correlations with the bisector and the contrast and other activity indicators are also found.
 Further relationships between the FWHM, and the BIS with the radial velocity variations are found
 in agreement with previous results based on solar-type stars \citep{2018A&A...616A.155L}.
 The disc-integrated magnetic field also is shown to correlate with the radial velocity variations
 as already found 
 \citep{2016MNRAS.457.3637H,2016A&A...587A.103L}.

 We finally focused on the temporal evolution of the $S$ index - H$\alpha$ correlation and the $S$ index - radial velocity correlation.
 We show how these correlations depend on the presence of inhomogeneities on the solar surface (as measured by sun spot numbers, irradiance,
 and radio flux values). 
 At larger levels of activity Ca and H$\alpha$ (and Ca and radial velocity variations)
 show a higher degree of correlation, 
 suggesting significant changes in the relative contributions of bright plages and dark filaments.

 Combining the results from this work with additional studies of solar observations 
 will help us to improve our understanding of stellar signals, 
 allowing us to develop  mitigation techniques that could be 
 used to confirm small planets  around low-mass stars.
 In particular, an in-depth analysis of solar images would  help us to unravel the correlation between 
 the S and H$\alpha$ indexes, and to better understand the behaviour of these lines in the minimum phase of the activity cycle.

\begin{acknowledgements}

 J.M., A.F.L., G.M., A.S., L.M., E.M., G.P., and E.P. acknowledge the support by INAF/Frontiera through the ``Progetti Premiali'' funding scheme
 of the Italian Ministry of Education, University, and Research.
 D.F.P.  is supported by NASA award number NNX16AD42G.
 X.D. is grateful to the Branco-Weiss Fellowship—Society in Science for financial support.
 A.C.C. acknowledges support from
 the Science and Technology Facilities Council (STFC) consolidated
 grant number ST/R000824/1.
 This work was performed in part under contract with the California Institute of Technology (Caltech)/Jet Propulsion Laboratory (JPL) funded by NASA through the Sagan Fellowship
  Program executed by the NASA Exoplanet Science Institute (R.D.H.)
 S.H.S. was supported by a NASA Heliophysics LWS grant NNX16AB79G.
 H.M.C. acknowledges support from the National Centre for Competence in Research (NCCR) ``PlanetS'' supported by the Swiss National Science Foundation (SNSF).
Based on observations made with the Italian Telescopio Nazionale Galileo (TNG), operated on the island of La Palma by the Fundaci\'on Galileo Galilei of the INAF (Istituto Nazionale di Astrofisica) at the Spanish Observatorio del Roque de los Muchachos of the Instituto de Astrof\'isica de Canarias.
The solar telescope used in these observations was built and maintained with support from the Smithsonian Astrophysical Observatory, the Harvard Origins of Life Initiative, and the TNG.
The HARPS-N project has been funded by the Prodex Program of the Swiss Space Office (SSO),
the Harvard University Origins of Life Initiative (HUOLI), the Scottish Universities Physics Alliance (SUPA),
the University of Geneva, the Smithsonian Astrophysical Observatory (SAO), the Italian National Astrophysical Institute (INAF), 
the University of St Andrews, Queen's University Belfast, and the University of Edinburgh.  
 Based on SOLIS data obtained by the NSO Integrated Synoptic Program (NISP), managed by the National Solar Observatory, which is operated by the Association of Universities for Research in Astronomy (AURA), Inc. under a cooperative agreement with the National Science Foundation,
 data obtained by the  WDC-SILSO, Royal Observatory of Belgium, Brussels,
 data obtained by the Solar Radiation and Climate Experiment (SORCE) satellite mission
 operated by the Laboratory for Atmospheric and Space Physics (LASP) at the University of Colorado (CU) in Boulder, Colorado, USA,
 and  Penticton radio fluxes data from NOAA's NGDC. 
 We sincerely appreciate the careful reading of the manuscript and the constructive comments of the anonymous referee.

\end{acknowledgements}

%
\bibliographystyle{aa}
\bibliography{harpn_sun.bib}


\begin{appendix}
\section{Sliding periodograms}

\begin{figure*}[!htb]
\centering
\begin{minipage}{0.48\linewidth}
\includegraphics[scale=0.55]{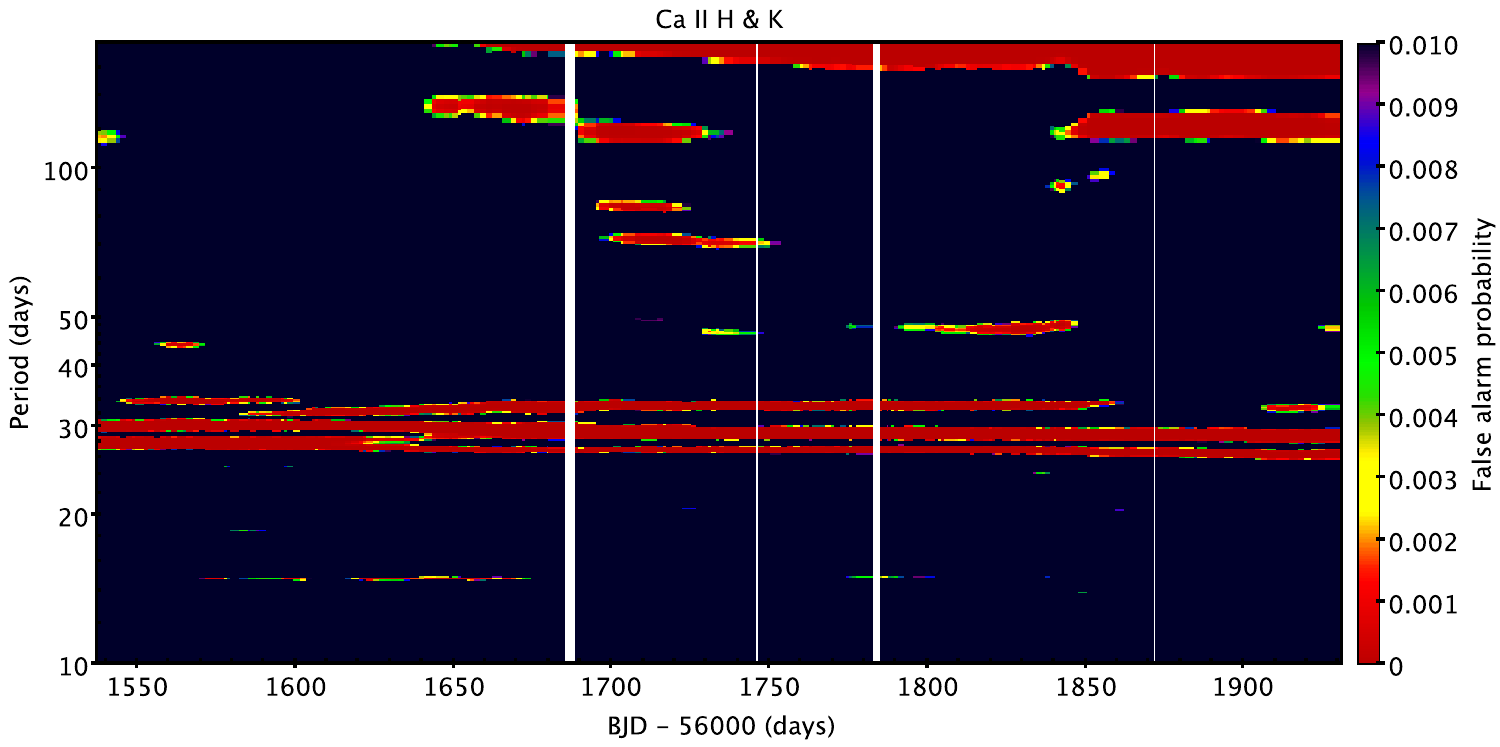} 
\end{minipage}
\begin{minipage}{0.48\linewidth}
\includegraphics[scale=0.55]{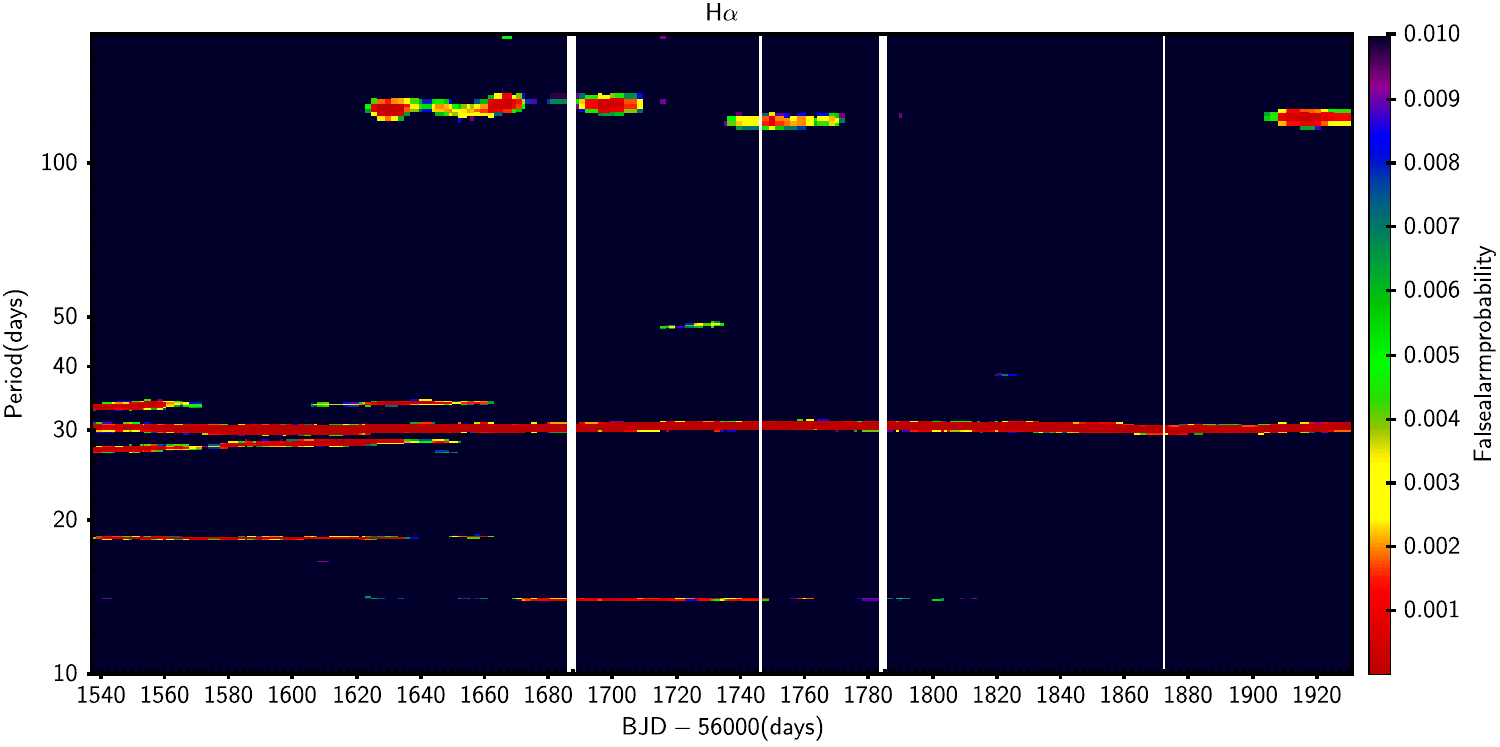}
\end{minipage}
\begin{minipage}{0.48\linewidth}
\includegraphics[scale=0.55]{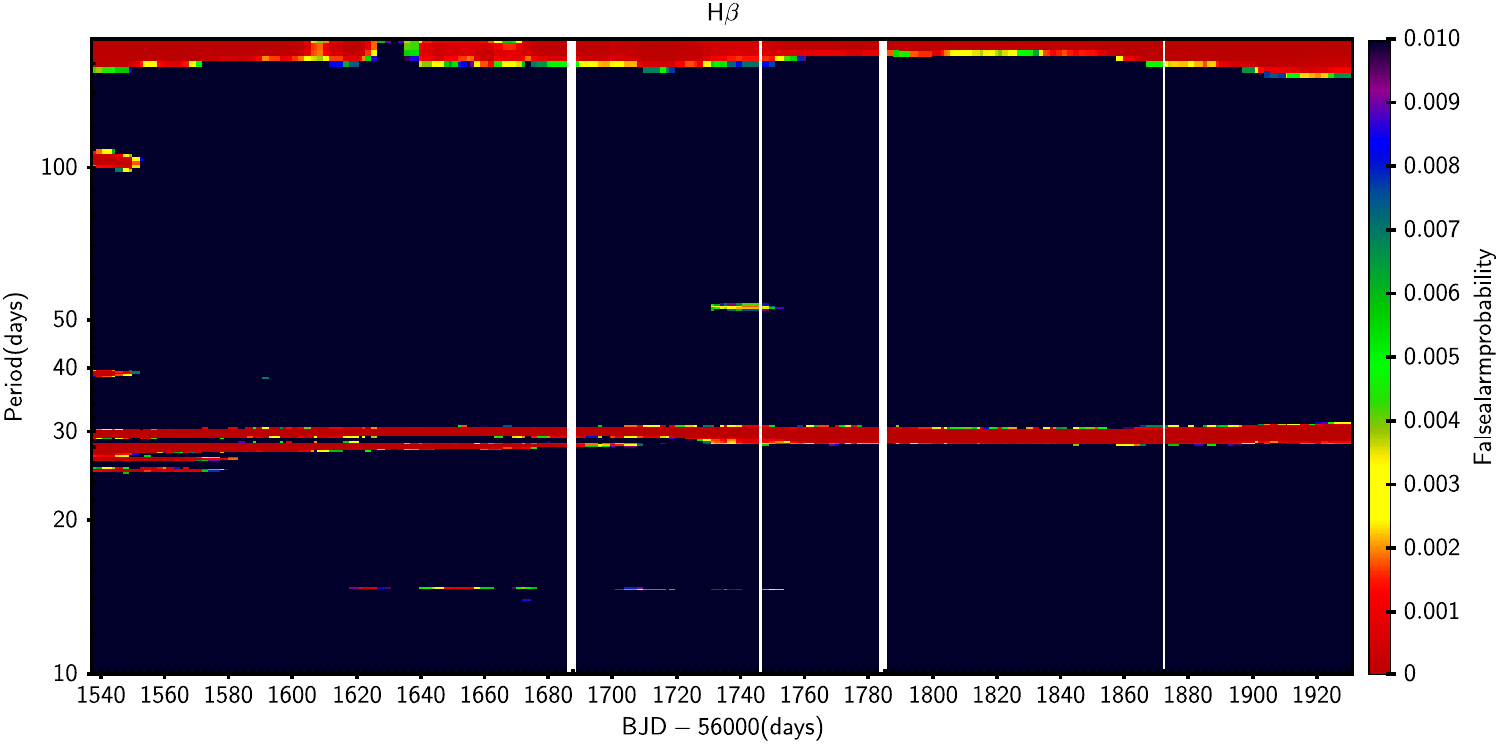}
\end{minipage}
\begin{minipage}{0.48\linewidth}
\includegraphics[scale=0.55]{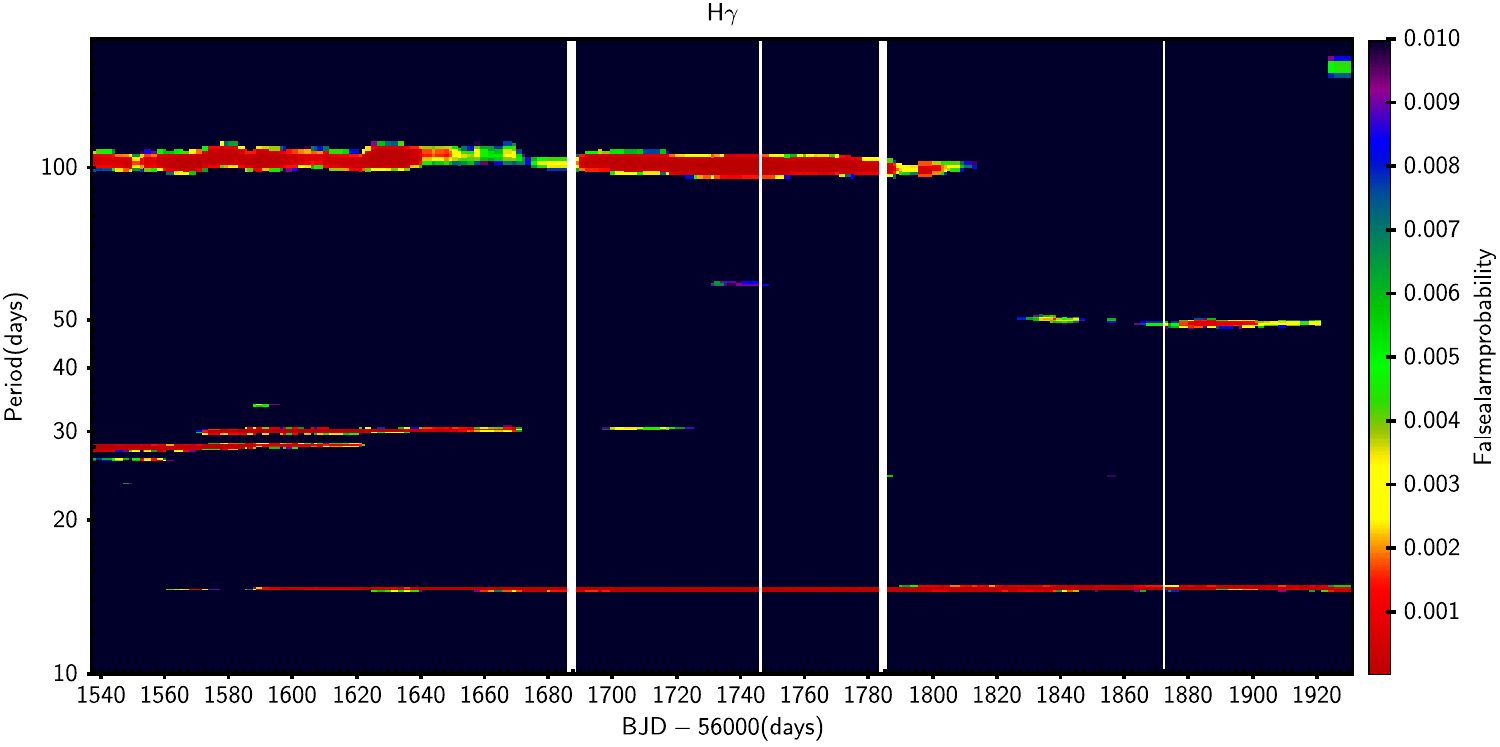}
\end{minipage}
\begin{minipage}{0.48\linewidth}
\includegraphics[scale=0.55]{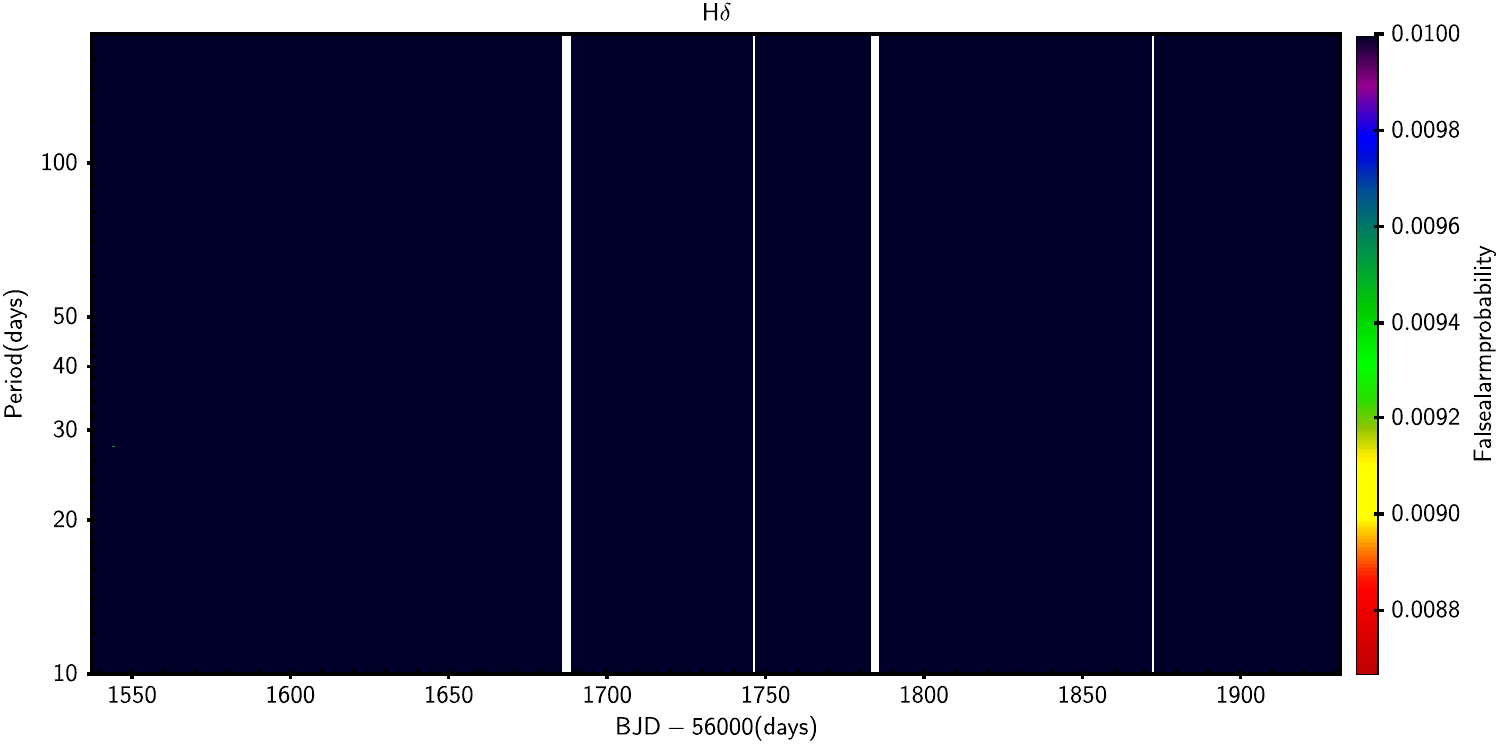}
\end{minipage}
\begin{minipage}{0.48\linewidth}
\includegraphics[scale=0.55]{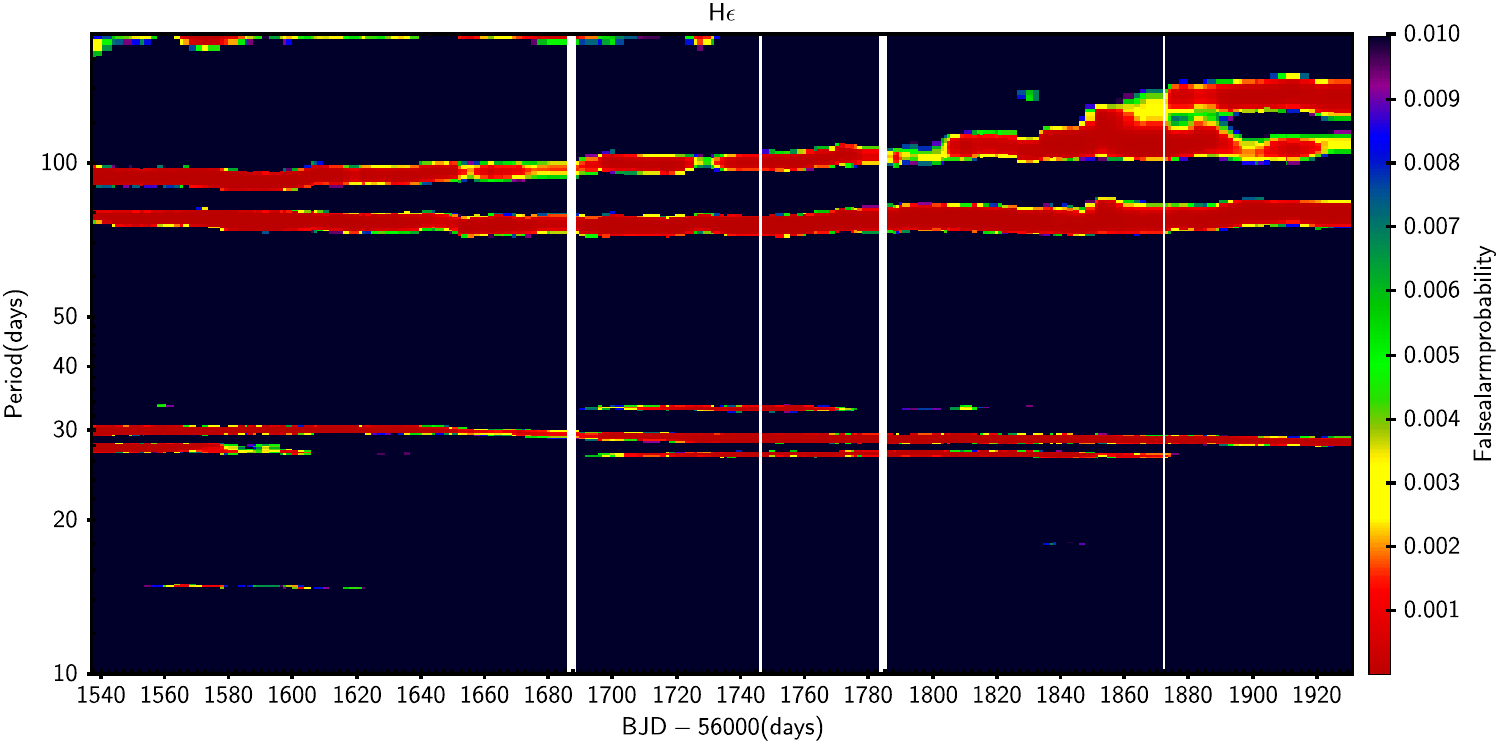}
\end{minipage}
\begin{minipage}{0.48\linewidth}
\includegraphics[scale=0.55]{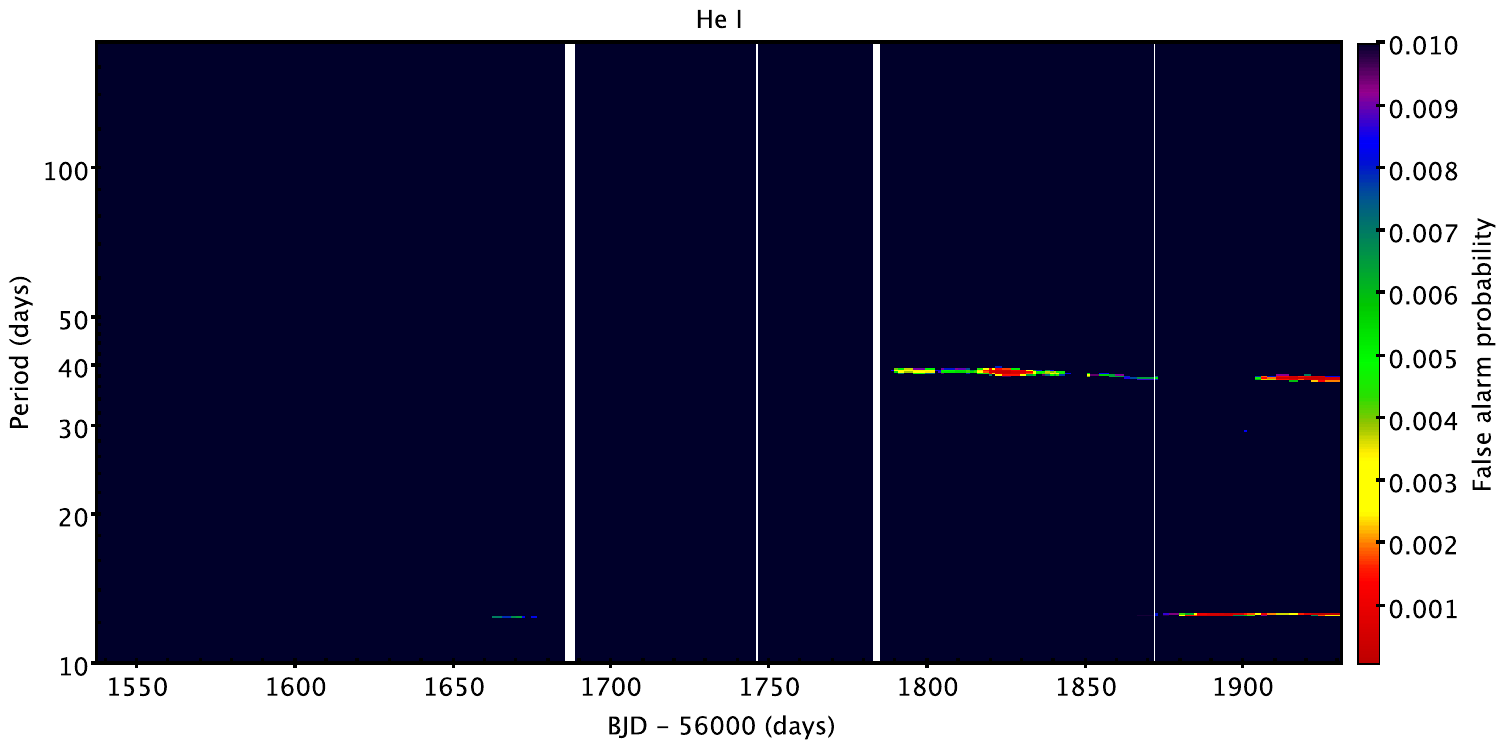}
\end{minipage}
\begin{minipage}{0.48\linewidth}
\includegraphics[scale=0.55]{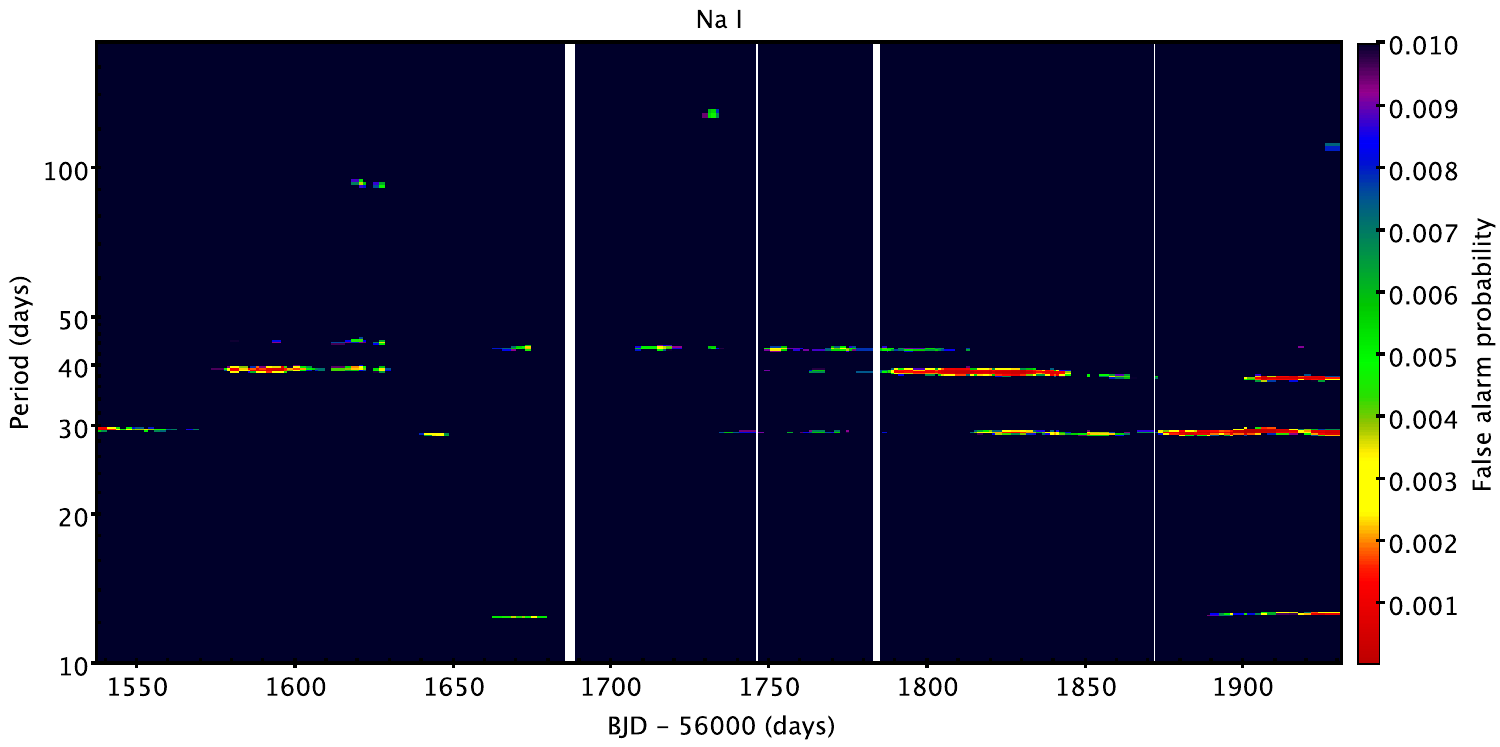}
\end{minipage}
\caption{ 
Sliding periodograms computed with a time window of 600 days.
From left to right and from top to bottom:
Ca~{\sc ii}, H$\alpha$, H$\beta$, H$\gamma$, H$\delta$, H$\epsilon$, He~{\sc i}, and Na~{\sc i}.
The X-axis shows the median time over which the GLS periodograms are calculated. Vertical white strips corresponds to gaps in the data.
}
\label{peridogramas_hbeta_900}
\end{figure*}

\begin{figure*}[!htb]
\centering
\includegraphics[scale=0.65]{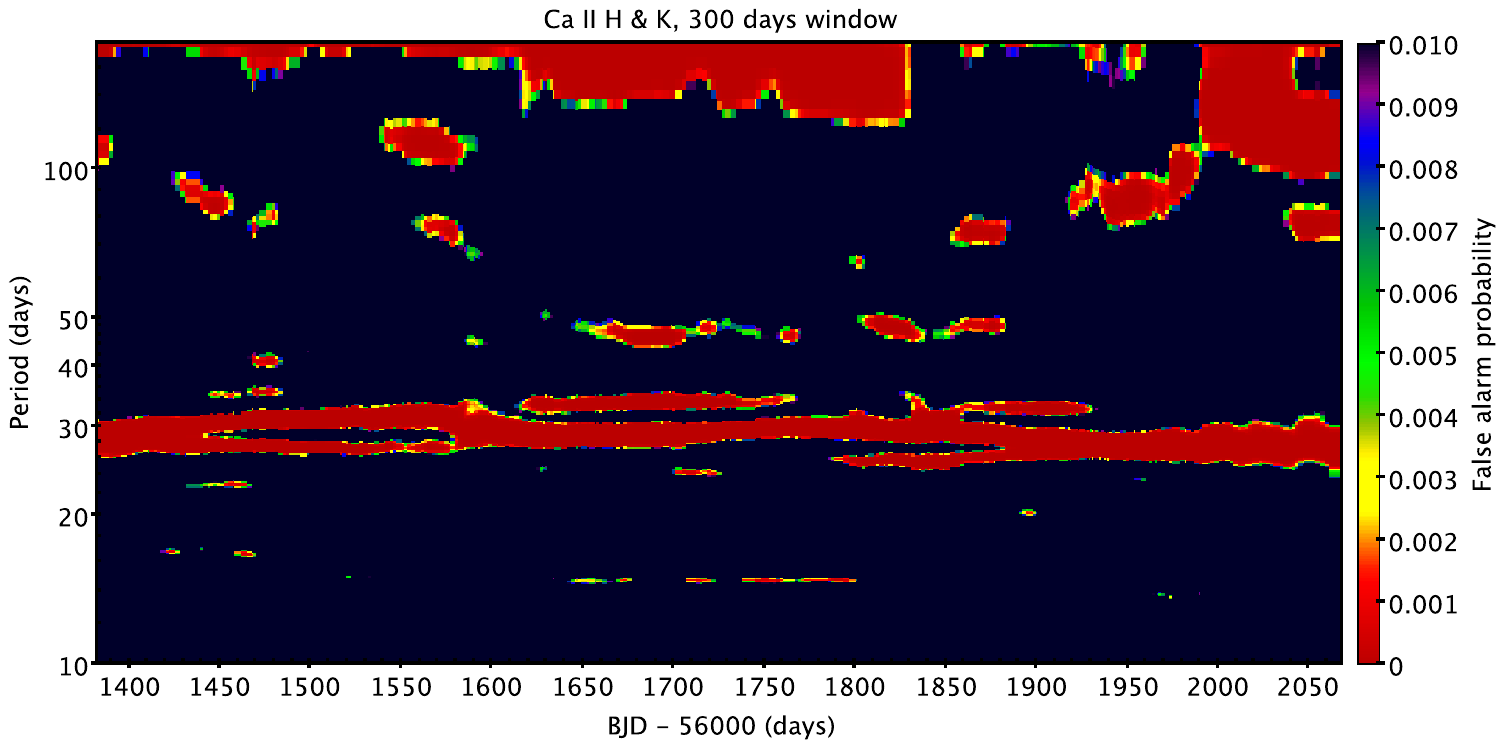}\\ 
\includegraphics[scale=0.65]{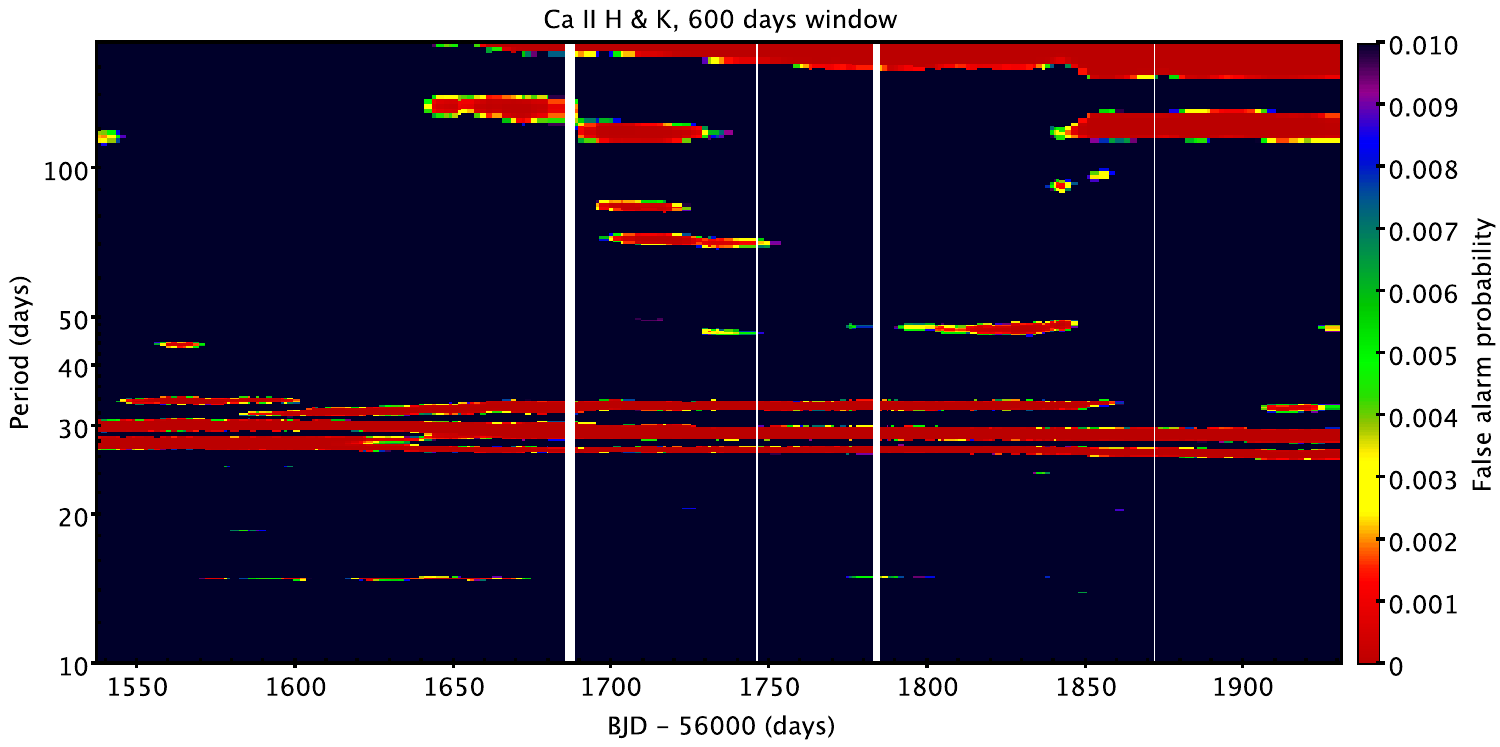}\\ 
\includegraphics[scale=0.65]{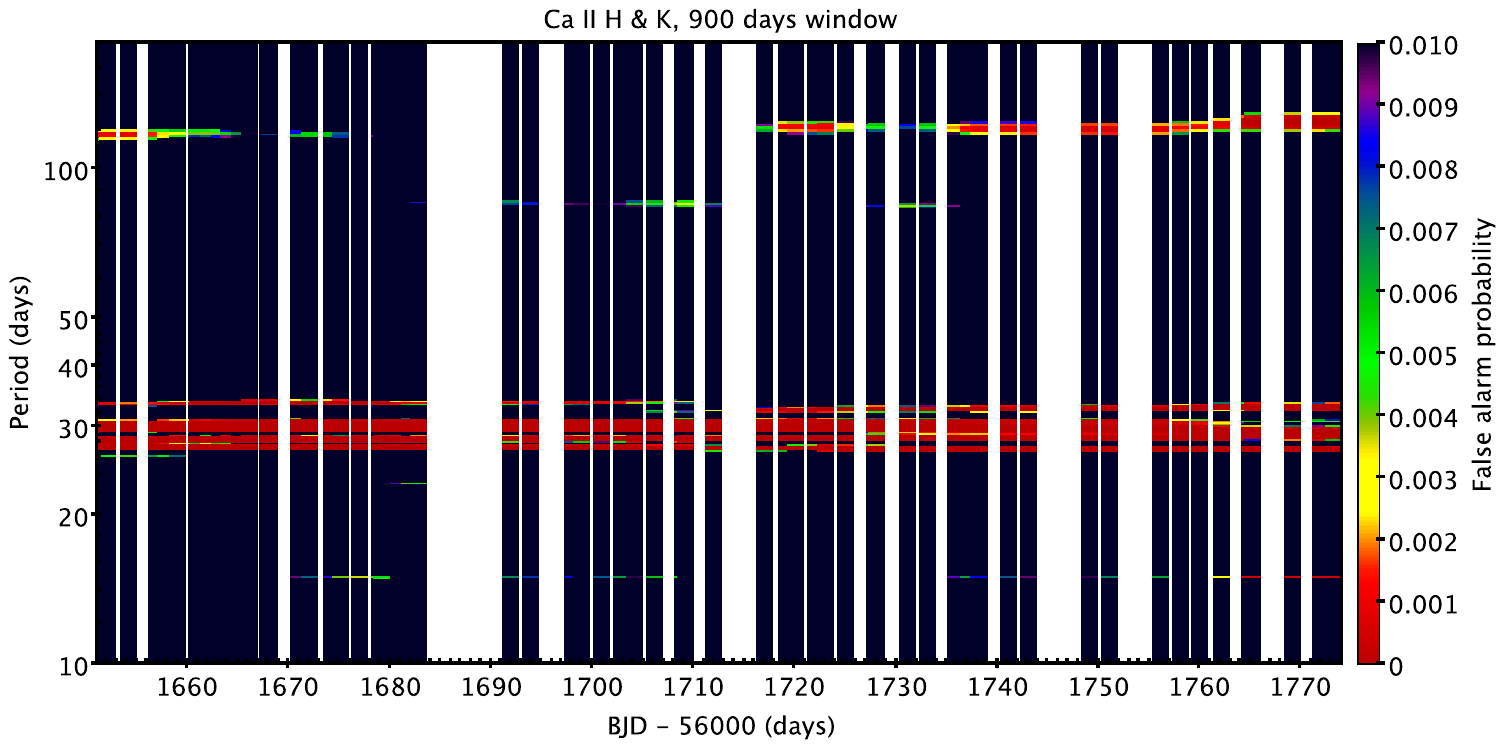}    
\caption{  Sliding periodograms for Ca~{\sc ii} computed with a time window of 300 (top), 600 (middle), and 900 days (bottom).
The X-axis shows the median time over which the GLS periodograms are calculated. Vertical white strips corresponds to gaps in the data.}
\label{peridogramas_window}
\end{figure*}

\begin{figure*}[!htb]
\centering
\includegraphics[scale=0.65]{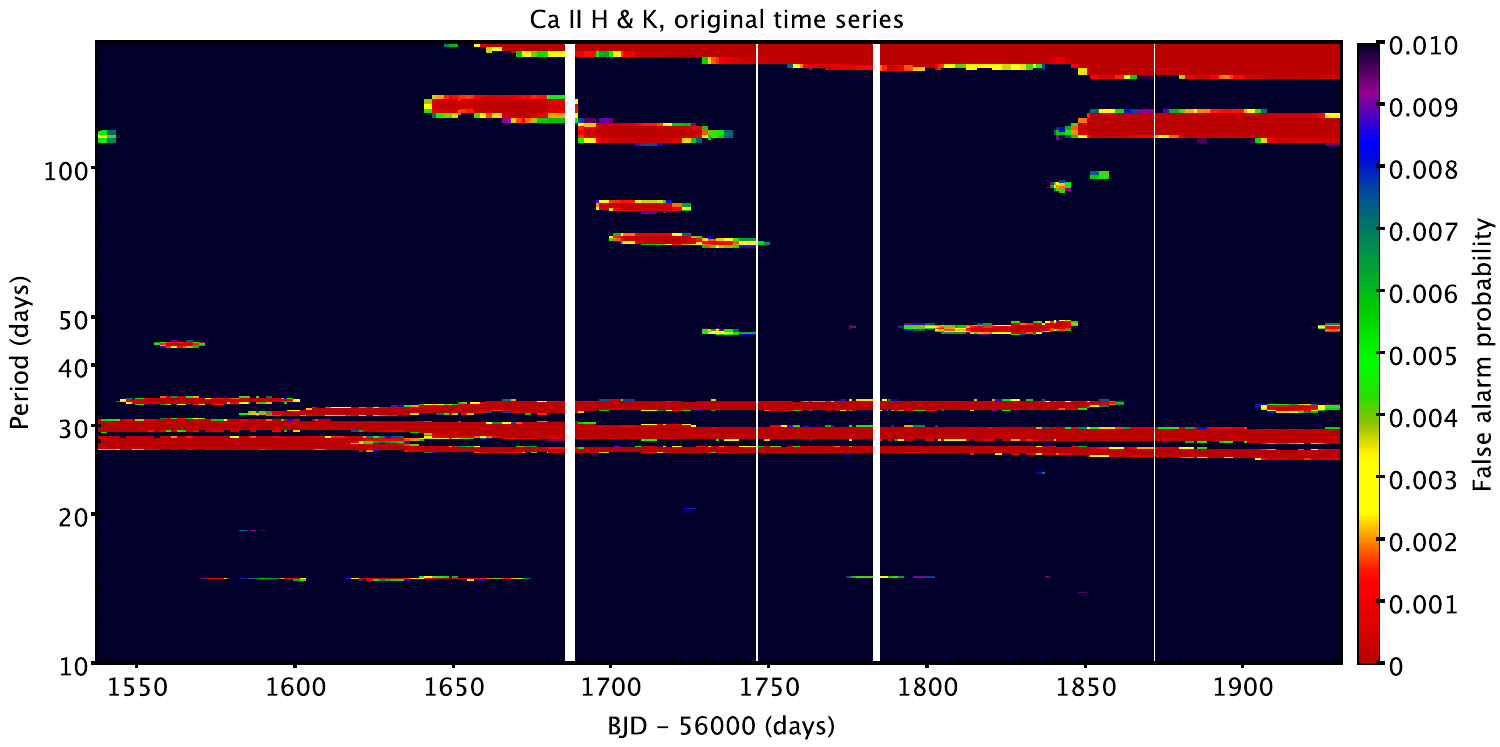}\\  
\includegraphics[scale=0.65]{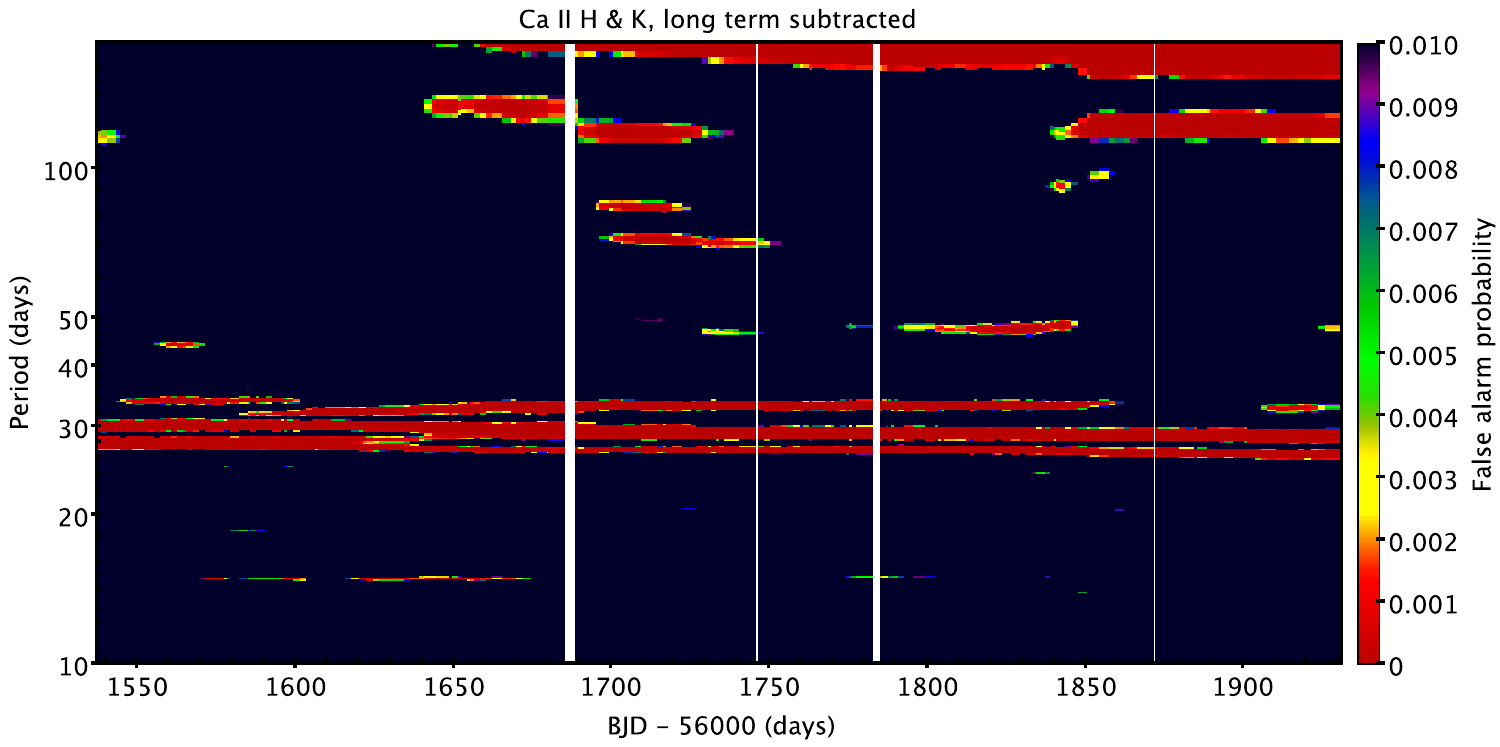} 
\caption{   Sliding periodograms for Ca~{\sc ii} computed
for the original time series (top), and after prewhitening of the long-term period (bottom). 
The temporal window has been fixed to 600 days in all cases.
The X-axis shows the median time over which the GLS periodograms are calculated. Vertical white strips corresponds to gaps in the data.}
\label{peridogramas_prewhitening}
\end{figure*}

\end{appendix}

\end{document}